\documentclass[sigconf,nonacm,screen]{acmart}
\pdfoutput=1

\title{THIA: Accelerating Video Analytics using\\ 
Early Inference and Fine-Grained Query Planning}

\usepackage{graphicx}
\usepackage{multirow}
\usepackage[ruled,vlined,linesnumbered]{algorithm2e}
\usepackage{subcaption}
\usepackage{booktabs}
\usepackage{pifont}
\usepackage{lipsum}
\usepackage{makecell}
\usepackage[para,online,flushleft]{threeparttable}
\usepackage{mathtools}
\usepackage{titlesec}
\usepackage{listings}
\usepackage{wrapfig}
\usepackage{balance}

\SetCommentSty{mycommfont}

\DeclarePairedDelimiter{\ceil}{\lceil}{\rceil}

\titlespacing*{\section}{0pt}{10pt minus 2pt}{3pt}
\titlespacing*{\subsection}{0pt}{5pt minus 2pt}{3pt}

\newcommand{\eg}{\textit{e.g.,}\xspace}
\newcommand{\ie}{\textit{i.e.,}\xspace}
\newcommand{\vs}{\textit{vs.}\xspace}

\newcommand{\rpn}{RPN\xspace}
\newcommand{\roi}{ROI\xspace}
\newcommand{\dnn}{DNN\xspace}

\newcommand{\fonescore}{F-1\xspace}

\newcommand{\X}{$\times$\xspace}

\newcommand{\chunk}{chunk\xspace}
\newcommand{\sampleover}{optimization time\xspace}
\newcommand{\execover}{execution time\xspace}

\newcommand{\qone}{\textsc{Q1}\xspace}
\newcommand{\qtwo}{\textsc{Q2}\xspace}
\newcommand{\qthree}{\textsc{Q3}\xspace}
\newcommand{\qfour}{\textsc{Q4}\xspace}

\newcommand{\ei}{\textsc{Early Inference}\xspace}
\newcommand{\ep}{EP\xspace}
\newcommand{\eps}{EPs\xspace}
\newcommand{\ds}{\textsc{Fine-Grained Planning}\xspace}
\newcommand{\me}{\textsc{Exit Point Estimation}\xspace}

\newcommand{\plan}{\textsc{Optimizer}\xspace}
\newcommand{\exec}{\textsc{Execution Engine}\xspace}

\newcommand{\ignore}[1] {}

\newcommand{\sys}{\textsc{Thia}\xspace}
\newcommand{\sysnaive}{\textsc{Naive}\xspace}
\newcommand{\syspp}{\textsc{PP}\xspace}
\newcommand{\sysblaze}{\textsc{BlazeIt}\xspace}
\newcommand{\syssingle}{\textsc{Thia-Single}\xspace}
\newcommand{\sysmulti}{\textsc{Thia-Multi}\xspace}
\newcommand{\sysei}{\textsc{Thia-EI}\xspace}

\newcommand{\PP}[1]{\vspace{4px}\noindent{\bf\textsc{#1}}\xspace}

\def\Snospace~{\S{}}

\newcommand\BeraMonottfamily{%
  \def\fvm@Scale{0.85}% scales the font down
  \fontfamily{fvm}\selectfont% selects the Bera Mono font
}

\lstdefinestyle{SQLStyle}{
  language=SQL,
  basicstyle=\BeraMonottfamily\footnotesize, 
  keywordstyle=\color{codepurple}\bfseries,
  showstringspaces=false,
  aboveskip = 0.05in,
  belowskip = 0.05in,
  literate = {-}{-}1, % <------ trick!
}

\makeatletter
\newcommand*{\rom}[1]{\expandafter\@slowromancap\romannumeral #1@}
\makeatother

\newcommand{\squishitemize}{
 \begin{list}{$\bullet$}
  { \setlength{\itemsep}{0pt}
     \setlength{\parsep}{3pt}
     \setlength{\topsep}{3pt}
     \setlength{\partopsep}{0pt}
     \setlength{\leftmargin}{1.95em}
     \setlength{\labelwidth}{1.5em}
     \setlength{\labelsep}{0.5em} } }

\newcounter{Lcount}
\newcommand{\squishlist}{
    \begin{list}{\arabic{Lcount}. }
   { \usecounter{Lcount}
        \setlength{\itemsep}{0pt}
        \setlength{\parsep}{3pt}
        \setlength{\topsep}{3pt}
        \setlength{\partopsep}{0pt}
        \setlength{\leftmargin}{2em}
        \setlength{\labelwidth}{1.5em}
        \setlength{\labelsep}{0.5em} } }

\newcommand{\squishend}{\end{list}}

\definecolor{linkcolor}{HTML}{647382}
\definecolor{citecolor}{HTML}{647382}
\definecolor{urlcolor}{rgb}{0.4,0.2,0.2}
\definecolor{sqlcolor}{HTML}{965d67}
\definecolor{smtcolor}{HTML}{5d968c}

\definecolor{codepurple}{HTML}{C42043}

\usepackage{hyperref}
\hypersetup{
    colorlinks=true,
    linkcolor=blue,
    filecolor=magenta,      
    urlcolor=cyan,
}

\usepackage[capitalise,noabbrev]{cleveref}

\begin{document}

%%%%%%%%%%%%%%%%%%%%%%%%%%%%%%%%%%%%%%%%%%%%%%%%%%%%%%%%%%%%%%%%%%%%%%%
% Author.
%%%%%%%%%%%%%%%%%%%%%%%%%%%%%%%%%%%%%%%%%%%%%%%%%%%%%%%%%%%%%%%%%%%%%%%
\author{Jiashen Cao}
\affiliation{%
  \institution{Georgia Institute of Technology}
}
\author{Ramyad Hadidi}
\affiliation{%
  \institution{Georgia Institute of Technology}
}
\author{Joy Arulraj}
\affiliation{%
  \institution{Georgia Institute of Technology}
}
\author{Hyesoon Kim}
\affiliation{%
  \institution{Georgia Institute of Technology}
}

%%%%%%%%%%%%%%%%%%%%%%%%%%%%%%%%%%%%%%%%%%%%%%%%%%%%%%%%%%%%%%%%%%%%%%%
% Author end.
%%%%%%%%%%%%%%%%%%%%%%%%%%%%%%%%%%%%%%%%%%%%%%%%%%%%%%%%%%%%%%%%%%%%%%%

% content
\begin{abstract}
To efficiently process visual data at scale, researchers have proposed
two techniques for lowering the computational overhead associated with the
underlying deep learning models. 
The first approach consists of leveraging a specialized, lightweight model
to directly answer the query.
The second approach focuses on filtering irrelevant frames using a lightweight
model and processing the filtered frames using a heavyweight model.
These techniques suffer from two limitations. 
With the first approach, the specialized model is unable to provide accurate
results for hard-to-detect events. 
With the second approach, the system is unable to accelerate queries focusing on
frequently occurring events as the filter is unable to eliminate a significant
fraction of frames in the video.

In this paper, we present \sys, a video analytics system for tackling these
limitations.
The design of \sys is centered around three techniques.
First, instead of using a cascade of models, it uses a single object detection
model with multiple exit points for short-circuiting the inference.
This early inference technique allows it to support a range of
throughput-accuracy tradeoffs.
Second, it adopts a fine-grained approach to planning, and processes different 
chunks of the video using different exit points to meet the user's requirements.
Lastly, it uses a lightweight technique for directly estimating the exit point
for a chunk to lower the optimization time. 
We empirically show that these techniques enable \sys to outperform two
state-of-the-art video analytics systems by up to 6.5\X, 
while providing accurate results even on queries focusing on hard-to-detect
events.
\end{abstract}

\maketitle

\section{Introduction}
\label{sec:intro}

%%%%%%%%%%%%%%%%%%%%%%%%%%%%%%%%%%%%%%%%%%%%%%%%%%%%%%%%%%%%%%%%%%%%%%%
% Discuss the why we need a video analytics system.
% And explain the challenge of building a video analytics system.
%%%%%%%%%%%%%%%%%%%%%%%%%%%%%%%%%%%%%%%%%%%%%%%%%%%%%%%%%%%%%%%%%%%%%%%
%
Researchers have proposed systems for quickly processing visual
data  with a tolerable drop
in accuracy~\cite{video-db1,video-db2,video-db3,noscope,blazeit,panorama,pp,focus,edge-va1,va1,va2}.
These systems detect objects in videos using deep neural networks
({\dnn}s)~\cite{faster-rcnn,mask-rcnn}.
%
%For example, a traffic regulator may use one of these systems to study the
%traffic patterns~\cite{traffic-va-1,traffic-va-2} in videos collected from
%surveillance cameras.
%
The key challenge that these systems tackle is the computational overhead
of the underlying object detection model.

%%%%%%%%%%%%%%%%%%%%%%%%%%%%%%%%%%%%%%%%%%%%%%%%%%%%%%%%%%%%%%%%%%%%%%%
% Introduce current video analytics system and discuss limitations.
%%%%%%%%%%%%%%%%%%%%%%%%%%%%%%%%%%%%%%%%%%%%%%%%%%%%%%%%%%%%%%%%%%%%%%%

\PP{Prior Work}.
To efficiently process visual data at scale, researchers have proposed two 
techniques.
%
%To avoid the cost of running the compute-intensive object detection model, 
%prior efforts~\cite{pp,blazeit} seek to skip 
%frames 
%
The first approach, presented in \sysblaze~\cite{blazeit}, consists of
leveraging a specialized, lightweight model to directly answer the query.
The second approach, introduced in \syspp~\cite{pp}, focuses on filtering
irrelevant frames using a lightweight model.
The frames that pass through the filtering model are then processed by the
heavyweight object detection model (illustrated in~\cref{fig:back:sys}). 
So, these systems accelerate query processing by \textit{not} processing a
subset of video frames using the heavyweight model.
However, these techniques suffer from two limitations.
With the first approach, the specialized model is unable to provide accurate
results for hard-to-detect events.
With the second approach, the system is unable to accelerate queries focusing on
frequently occurring events.
This is because the filter is unable to eliminate a significant fraction of
frames in the video.

Another line of research, illustrated in \textsc{Tahoma}~\cite{tahoma}, focuses
on leveraging a collection of differently sized models to process the frames
based on the complexity of the event.
However, using such a cascade of models comes with two limitations.
First, switching from one model to another in the GPU is expensive.
This switching overhead is further exacerbated if we seek to frequently change
the model to process different subsets (\ie \chunk{s}) of the video 
to maximize performance.
For instance, loading a Faster-RCNN model in PyTorch on an NVIDIA 
Titan Xp GPU takes 2~s 
(including framework initialization and model loading).
Second, using a collection of models to support different throughput-accuracy 
tradeoffs does not scale well due to the large GPU memory footprint of these
models.
%
%As shown in~\cref{fig:intro:mem},
%the memory usage of using separate models significantly
%increases as the number of supported 
%\textit{exit points}\footnote{We refer different accuracy and 
%speedup trade-off points in object detection model as exit points.} increases. 

Prior efforts have mostly focused on altering the design of the inference
pipeline.
However, they do not elaborate on how to adapt this pipeline 
(\eg when to use a particular model) based on the \chunk.
They choose a single plan for the entire video based on the profiling results
obtained on a set of sampled frames.
Such a coarse-grained approach to query planning does not leverage the variation
in the frequency and detection difficulty of different events in a video.
If objects are difficult to detect, this approach leads to less accurate
results.
On the other hand, if objects are easier to detect, a conservative
coarse-grained query plan significantly increases the query processing time
(but returns correct results).
We defer a detailed discussion of these limitations to~\autoref{sec:mov}.

\PP{Our Approach}.
In this paper, we present \sys, a video analytics system for tackling the
limitations highlighted above.
\sys leverages three techniques to accelerate queries over visual data.

First, it uses a \textit{single} object detection model with multiple points for
short-circuiting the inference.
These \textit{exit points} (\eps) offer a set of throughput-accuracy tradeoffs.
While processing the query, \sys uses a shallow \ep to quickly process frames
that are irrelevant or contain easy-to-detect events.
If the frames contain hard-to-detect events, then \sys falls back to a deeper
\ep in the model to deliver higher accuracy.
This \ei technique eliminates the switching overhead and lowers the
GPU memory footprint of \sys
%\footnote{
%The \ei technique is orthogonal to the model ensemble technique.
%
%It can be applied for each model in a collection of models to obtain a broader
%set of accuracy-performance tradeoffs.
%}
.
%
%
% It offers more flexibility of constructing knobs than using multiple
% models.
% 
% The \ei technique is able to support more knobs with the same 
% model parameter constraints.
%
% For example, \ei technique offers five knobs with 1GB model parameters.
%
% Instead, with 1GB constraint, we can only support three models (three knobs) if 
% using multiple models.
%
% Because all knobs constructed by the \ei technique share common layers, 
% they can be trained together in one pass.
%
% In contrast, knobs of multiple models approach have to be trained separately.
%
% Second, from the system perspective, users have to change object detection model 
% architecture manually to obtain models with different scales in the
% case of using multiple models.
%
% \htodo{do you have a comparison of the number of parameters for different approaches?%
% Probably good to show it.}%
%
% \htodo{probably better to define what knob means in this context more clearly?}%
% \htodo{this seems to be always good feature to have.%
% can you elaborate more why others can't do that?  but EI can do that?}%

Second, \sys adopts a fine-grained approach to planning.
It processes different chunks of the video using different \eps to meet both the
performance and accuracy requirements (elaborated in~\autoref{sec:plan:need}).
This \ds technique increases the \sampleover of a query.
To lower this overhead, we present a third technique to quickly decide which
\ep to use for a given chunk.
\sys uses a shallow model for \me instead of running inference on the sampled
frames.
%
%%%%%%%%%%%%%%%%%%%%%%%%%%%%%%%%%%%%%%%%%%%%%%%%%%%%%%%%%%%%%%%%%%%%%%%
% Discuss experiments value and give big picture of our results.
%%%%%%%%%%%%%%%%%%%%%%%%%%%%%%%%%%%%%%%%%%%%%%%%%%%%%%%%%%%%%%%%%%%%%%%
%
We evaluate a set of queries focusing on events with different levels of
frequency and detection difficulty on two traffic surveillance datasets:
UA-DeTrac~\cite{ua_detrac} and Jackson Town~\cite{blazeit}.
On all of the queries, \sys outperforms the state-of-the-art video analytics
systems by up to 6.5\X with a tolerable drop in accuracy.
%

%%%%%%%%%%%%%%%%%%%%%%%%%%%%%%%%%%%%%%%%%%%%%%%%%%%%%%%%%%%%%%%%%%%%%%%
% List the contribution of this paper.
%%%%%%%%%%%%%%%%%%%%%%%%%%%%%%%%%%%%%%%%%%%%%%%%%%%%%%%%%%%%%%%%%%%%%%%
\PP{Contributions}.
Our research makes the following contributions:
\squishitemize
  \item We present the \ei technique to construct a single model that
  offers a set of throughput-accuracy tradeoffs for challenging vision tasks
  like object detection.
  \item We propose a \ds technique that works in tandem with the \ei technique.
  \item We present the \me technique to reduce the optimization overhead of the
  \ds technique.
  \item We implement all of these techniques in \sys and show that it
  outperforms two state-of-the-art video analytics systems on a wide range of
  queries. 
\squishend

%%%%%%%%%%%%%%%%%%%%%%%%%%%%%%%%%%%%%%%%%%%%%%%%%%%%%%%%%%%%%%%%%%%%%%%
% Provide some necessary information here for readers.
%%%%%%%%%%%%%%%%%%%%%%%%%%%%%%%%%%%%%%%%%%%%%%%%%%%%%%%%%%%%%%%%%%%%%%%
\section{Background}\label{sec:back}
We now present an overview of object detection and sampling 
techniques used in video analytics systems (\autoref{sec:back:obj_det} and
\autoref{sec:back:sample}).
We later discuss the key techniques used in state-of-the-art
systems (\autoref{sec:back:sys}).

%%%%%%%%%%%%%%%%%%%%%%%%%%%%%%%%%%%%%%%%%%%%%%%%%%%%%%%%%%%%%%%%%%%%%%%
% Go over object detection, especially backbone, RPN, ROI, which are 
% needed for later early inferencing technique introduction.
%%%%%%%%%%%%%%%%%%%%%%%%%%%%%%%%%%%%%%%%%%%%%%%%%%%%%%%%%%%%%%%%%%%%%%%
\subsection{Object Detection}\label{sec:back:obj_det}

\begin{figure}[t]
  \centering
  \includegraphics[width=0.85\linewidth]{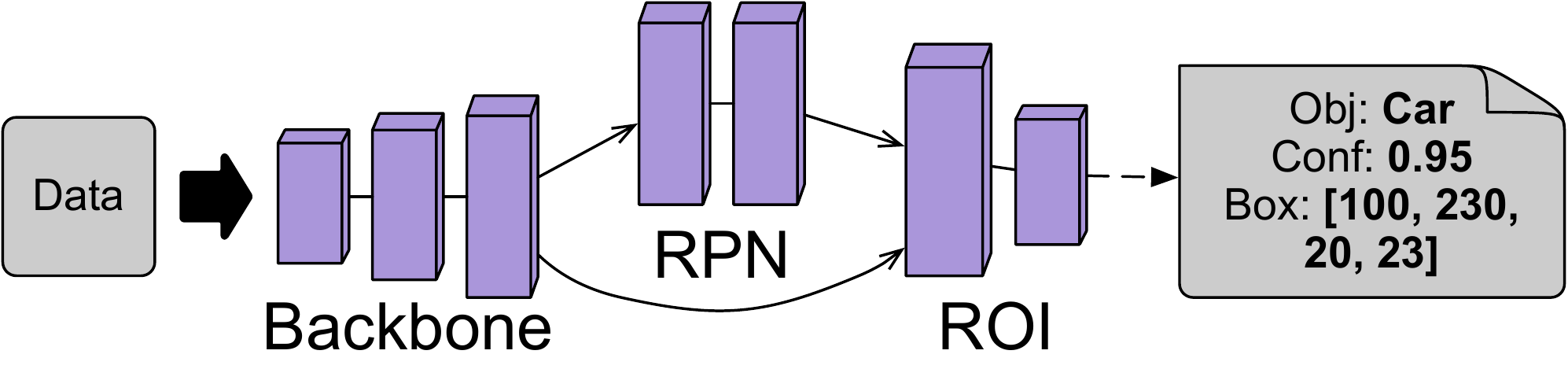}
  \caption{\textbf{Object Detection Model} -- Components of the 
  model.
  }
  \label{fig:back:obj_det}
\end{figure}

Object detection models usually contain three components: 
(1) backbone network, 
(2) region proposal network (\rpn), and 
(3) region of interests network (\roi),
as illustrated in~\cref{fig:back:obj_det}.
The backbone network extracts the high-level features from a frame. 
Then, the \rpn and \roi networks determine the location and type of objects
detected in the frame.
Data flows from the backbone network to \rpn, and \roi returns the final
prediction results (object category, location within the frame, and confidence
score).

In machine learning literature, the \textit{oracle} model returns the correct
answer to all queries. 
However, in practice, there is no ground truth for unseen data.
Similar to prior efforts, we assume that the most accurate model, which also
tends to be the most compute-intensive model, is the oracle 
model~\cite{blazeit,noscope,chameleon,pp,panorama}.
%

%%%%%%%%%%%%%%%%%%%%%%%%%%%%%%%%%%%%%%%%%%%%%%%%%%%%%%%%%%%%%%%%%%%%%%%
% Briefly discuss sampling here. The objective is to first define what
% is chunk, which will be used for query planning section. In addition,
% differentiate sampling and execution phase in our system, so readers
% understand why model estimation is needed for reducing the sampling
% overhead.
%%%%%%%%%%%%%%%%%%%%%%%%%%%%%%%%%%%%%%%%%%%%%%%%%%%%%%%%%%%%%%%%%%%%%%%
\subsection{Sampling}\label{sec:back:sample}
Sampling is a frequently used technique for processing visual data at scale.
By processing only a subset of frames using the object detection model, a video
analytics system lowers the overall query processing time.
For example, \sysblaze~\cite{blazeit} uses uniformly random sampling to process
aggregate queries (\eg counting the average number 
of cars within a given period of time).
%
%In this work, we only use sampling during the query planning.
%
%We extend the application of sampling technique to select query described 
%in~\cref{mov:qone}.

\sys uses sampling for a different purpose (elaborated in~\autoref{sec:plan}).
A \chunk is a continuous segment of frames within a video. 
\sys{'s} \plan constructs plans at \chunk-level granularity (instead of
video-level granularity) to lower the query processing time.
Query processing time consists of two components:
(1) \sampleover, and
(2) \execover.
During the optimization phase, \sys generates a plan for each \chunk.
During the execution phase, it runs these plans.

%%%%%%%%%%%%%%%%%%%%%%%%%%%%%%%%%%%%%%%%%%%%%%%%%%%%%%%%%%%%%%%%%%%%%%%
% Introduce four papers here: Panorama, Miris, PP, and BlazeIt. I want
% to first differentiate Panorama and Miris with our systems. Then 
% introduce the concept of PP and BlazeIt, which will be used for later
% motivation section to discuss their limitations.
%%%%%%%%%%%%%%%%%%%%%%%%%%%%%%%%%%%%%%%%%%%%%%%%%%%%%%%%%%%%%%%%%%%%%%%
\subsection{State-of-the-Art Systems}\label{sec:back:sys}

\begin{figure}[t]
  \centering
  \includegraphics[width=0.9\linewidth]{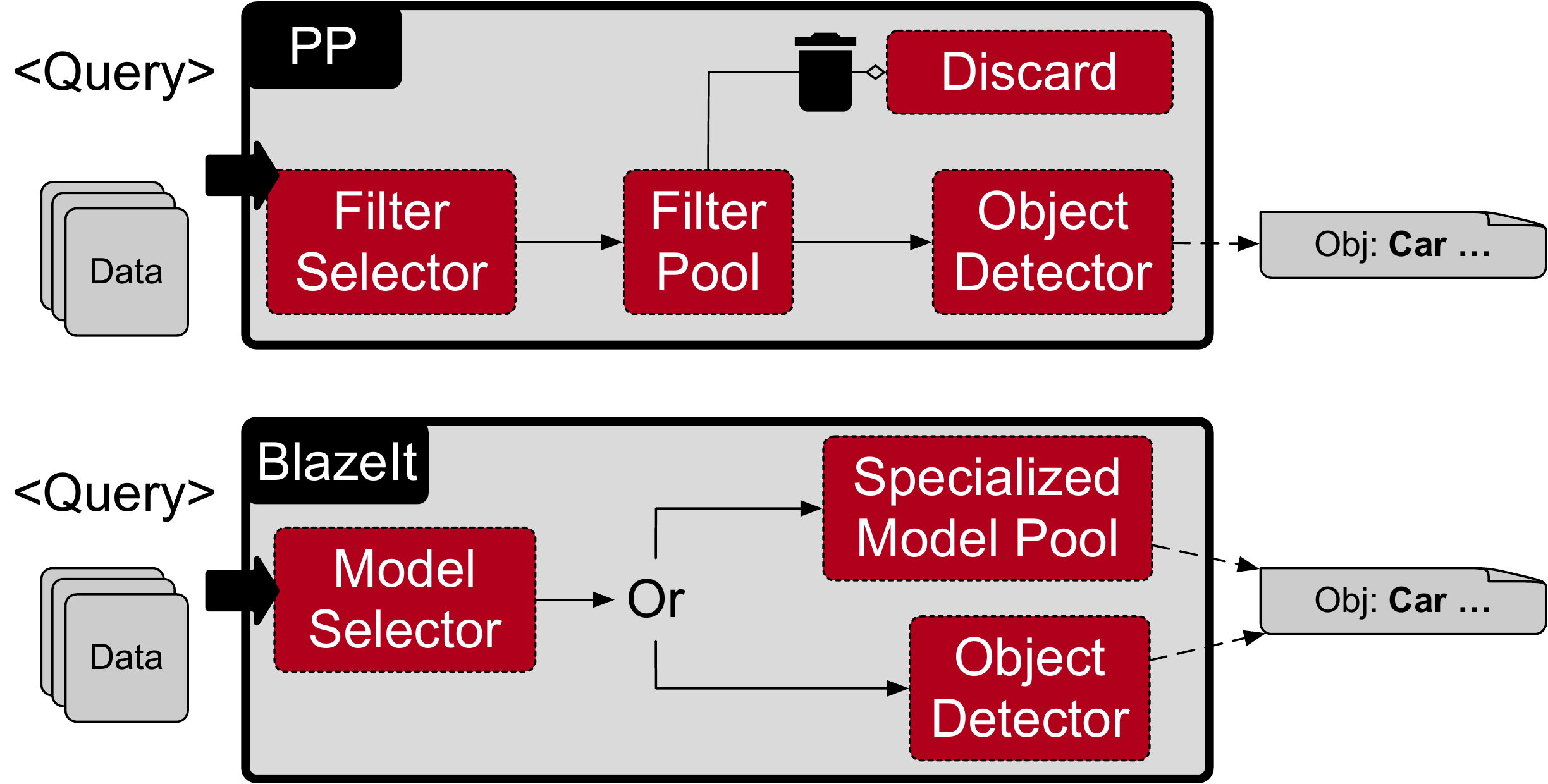}
  \caption{\textbf{Architecture of Video Analytics Systems} -- 
  Architecture of two state-of-the-art video analytics systems: 
  (1) \syspp~\cite{pp} and (2) \sysblaze~\cite{blazeit}.}
  \label{fig:back:sys}
\end{figure}

\begin{table}[t]

  \small
  
  \begin{threeparttable}
  
  \renewcommand{\arraystretch}{1.1}
  \centering
  \begin{tabular}{@{}lccccc@{}}
  
  \toprule
  
  & \small{\textbf{+ MS}}
  & \small{\textbf{+ MC}}
  & \small{\textbf{+ FP}} \textsuperscript{\small{\ding{61}}}
  & \small{\textbf{+ EI}} \textsuperscript{\small{\ding{61}}}
  & \small{\textbf{+ EP-Est}} \textsuperscript{\small{\ding{61}}}\\
  
  \midrule
  
  \sysnaive
  &
  &
  &
  & \\
  
%  Panorama
%  & 
%  &
%  & 
%  & \small{\ding{52}} 
%  & \\
%  
%  Miris
%  &
%  &
%  & \small{\ding{52}} 
%  & 
%  & \\
%  
%  Tahoma
%  & 
%  & \small{\ding{52}} 
%  & 
%  &
%  & \\
  
  \syspp
  & \small{\ding{52}} 
  &
  &
  &
  & \\
  
  \sysblaze
  & \small{\ding{52}} 
  &
  &
  &
  & \\
  
  \midrule
  
  \syssingle\textsuperscript{\tiny{\ding{72}}}
  &
  &
  & \small{\ding{52}} 
  &
  & \\
  
  \sysmulti\textsuperscript{\tiny{\ding{72}}}
  &
  & \small{\ding{52}} 
  & \small{\ding{52}} 
  &
  & \\
  
  \sysei\textsuperscript{\tiny{\ding{72}}}
  &
  &
  & \small{\ding{52}} 
  & \small{\ding{52}} 
  & \\
  
  \sys\textsuperscript{\tiny{\ding{72}}}
  &
  &
  & \small{\ding{52}} 
  & \small{\ding{52}} 
  & \small{\ding{52}} \\
  
  \bottomrule
  
  \end{tabular}
  
  \begin{tablenotes}
    \item \textbf{MS}: Model Specialization, 
    \textbf{MC}: Model Cascade, 
    \textbf{FP}: \ds, 
    \textbf{EI}: \ei, 
    \textbf{EP-Est}: \me.
    \item \ding{61}: Techniques used in \sys.
    \item \ding{72}: Variants of \sys.
  \end{tablenotes}
  
  \end{threeparttable}

  \caption{\textbf{Qualitative Comparison of Video Analytics Systems} --
  Key characteristics of state-of-the-art video analytics systems.
  }
  \label{tb:back:sys_variants}
\end{table}

\cref{tb:back:sys_variants} lists the key characteristics of
several state-of-the-art video analytics systems: 
(1) \syspp~\cite{pp},
(2) \sysblaze~\cite{blazeit} 
(3) Miris~\cite{miris}, 
(4) Tahoma~\cite{tahoma}, and 
(5) Panorama~\cite{panorama}.
We present the benefits and limitations of the first two systems
in~\autoref{sec:intro}.
Their architectures are illustrated in~\cref{fig:back:sys}.

Miris~\cite{miris} is a video analytics system that focuses on multi-object
tracking.
It uses coarse-grained sampling to gain a high-level perspective of the video
and then gradually increases the sampling rate to improve the accuracy of
tracking.
\sys differs from Miris in two ways.
First, it is tailored for object detection.
Second, it only samples for query planning (not for query execution).
In~\autoref{sec:eval:end2end}, we illustrate the benefits of other optimizations
in \sys by comparing it against a variant of \sys that only uses \ds
(\syssingle in~\cref{tb:back:sys_variants}). 

\textsc{Tahoma}~\cite{tahoma} is another closely related analytics system.
It constructs a model cascade by combining a chain of image classification
models and determines when to short-circuit the inference based on 
the confidence score of prediction of each model.
Unlike \textsc{Tahoma}, \sys is geared toward object detection.
So, the inference result consists of a set of confidence scores for all the
objects present in the frame.
It is challenging to short-circuit the inference pipeline based on a set of
confidence scores.
In \sys, the \ei technique is guided by query accuracy (not model accuracy).
In~\autoref{sec:eval:end2end} and~\autoref{sec:eval:mcopt}, 
we illustrate the limitations of using a model
cascade by comparing \sys against a variant of \sys that uses \ds along with a
model cascade (\sysmulti in~\cref{tb:back:sys_variants}), instead of a single
\ei model 
\footnote{
\sysmulti delivers better performance than a naive model cascade due to the 
\ds technique.
With a naive model cascade, the system cannot directly process frames with the
optimal model.
It must use all the smaller models before stopping the inference at the optimal
model.
}.

Panorama~\cite{panorama} is another state-of-the-art video video analytics
system that uses a single model to solve the unbounded vocabulary problem in
object recognition.
While this system also offers a set of throughput-accuracy tradeoffs similar to
\sys, it is geared towards comparing embeddings from two input frames.
So, it selects the \ep based on the delta between two embeddings while
extracting the embeddings.
Lastly, it clusters these embeddings to recognize the objects in the input
frames.
In contrast, \sys uses \ei in the object detection model itself and 
seeks to reduce \sampleover using the \me technique.

%
% \syspp and \sysblaze are two systems that target at the general object detection
% problem as our system.
% %
% As shown in~\cref{fig:back:sys}, they adopt different system designs respectively.
% %
% As indicated in~\cref{tb:back:sys_variants}, both of them use \textsc{Model Specialization} technique.
% %
% In other words, other than the object detection model, both of those systems maintain 
% a set of faster but less accurate models, which include support vector machine and shallow
% \dnn. 
% %
% The \syspp system relies on the idea of filtering to achieve performance speedup. 
% %
% When the system gets a query, the selected filter decides whether the interested 
% object is present in the image.
% %
% If the interested object is present, then the object detection model 
% runs inference on the image to provide final results.
% %
% Otherwise, the image is filtered and discarded.
% %
% On the other hand, the \sysblaze picks the best specialized model and the model directly 
% provides answer to the query without the object detection model.
% %
% Both \syspp and \sysblaze require multiple models for different queries.
% %
% As indicated by~\cref{tb:back:sys_variants}, \sys with all our proposed techniques
% is our final system design.
% %
% We also introduce \sysei to help us examine the benefit of each technique.

%%%%%%%%%%%%%%%%%%%%%%%%%%%%%%%%%%%%%%%%%%%%%%%%%%%%%%%%%%%%%%%%%%%%%%%
% Discuss why our system is needed. First introduce the queries we will
% use. Then discuss the limitation of current approaches.
%%%%%%%%%%%%%%%%%%%%%%%%%%%%%%%%%%%%%%%%%%%%%%%%%%%%%%%%%%%%%%%%%%%%%%%
\section{Motivation}
\label{sec:mov}

In this section, we discuss the limitations of \syspp and \sysblaze to 
motivate the need for \sys.
We focus on the four queries described in~\cref{tb:mov:queries}.
These queries differ in:
(1) frequency of appearance of target objects in the video, and
(2) level of difficulty in providing a correct answer to a query.

\begin{table}[t]

  \small
  
  \begin{threeparttable}
  
  \renewcommand{\arraystretch}{1.3}
  \centering
  \begin{tabular}{@{}l@{}lcc@{}}
  
  \toprule
  
  \multirow{2}{*}{\textbf{Query}}
  & \multicolumn{1}{c}{\multirow{2}{*}{\textbf{SQL}}}
  & \multirow{2}{*}{\textbf{\makecell{Predicate\\Frequency}}}
  & \multirow{2}{*}{\textbf{\makecell{Predicate\\Difficulty}}} \\ \\
  
  \midrule
  
  \textbf{\qone} \label{mov:qone}
  & \begin{lstlisting}[style=SQLStyle]
      Select frameID 
      From UA-DeTrac 
      Where Count(Car) >= 4;
    \end{lstlisting}
  & Frequent
  & Easy \\
    
  \textbf{\qtwo} \label{mov:qtwo}
  & \begin{lstlisting}[style=SQLStyle]
      Select frameID 
      From UA-DeTrac 
      Where Count(Truck) >= 1;
    \end{lstlisting}
  & Frequent
  & Hard \\
    
  \textbf{\qthree} \label{mov:qthree}
  & \begin{lstlisting}[style=SQLStyle]
      Select frameID 
      From UA-DeTrac 
      Where Count(Bus) >= 4;
    \end{lstlisting}
  & Rare
  & Hard \\
    
  \textbf{\qfour} \label{mov:qfour}
  & \begin{lstlisting}[style=SQLStyle]
      Select frameID 
      From Jackson-Town 
      Where Count(Car) >= 4;
    \end{lstlisting}
  & Rare
  & Hard \\
  
  \bottomrule
  
  \end{tabular}
  
  \end{threeparttable}
  \caption{\textbf{List of Queries} -- Queries with varying frequency and levels of difficulty in detecting events.}
  \label{tb:mov:queries}
\end{table}

%%%%%%%%%%%%%%%%%%%%%%%%%%%%%%%%%%%%%%%%%%%%%%%%%%%%%%%%%%%%%%%%%%%%%%%
% Discuss model specialization. The main idea is engineers have to 
% maintain different models, so having multiple accuracy and speed 
% tradeoff in a single model is a better approach.
%%%%%%%%%%%%%%%%%%%%%%%%%%%%%%%%%%%%%%%%%%%%%%%%%%%%%%%%%%%%%%%%%%%%%%%
\PP{Limitation \rom{1} -- model specialization overhead}. 
Both \syspp and \sysblaze rely on specialized models.
~\syspp uses a specialized model as a filter.
Since each filter detects only one object category, 
it needs to train multiple lightweight models (\ie filters) during runtime 
to support different object categories.
\sysblaze uses a specialized model to directly return the results.
A model may directly return the count of cars in an image, 
so it must maintain multiple models for different predicates 
(\eg \lstinline[style=SQLStyle]{Count(Car)} is a predicate).
With this model specialization technique, these systems need to train and
maintain models for different objects and predicates, respectively.
We seek to reduce model maintenance overhead by offering a range of accuracy  
and query execution time tradeoffs in a single model.

%%%%%%%%%%%%%%%%%%%%%%%%%%%%%%%%%%%%%%%%%%%%%%%%%%%%%%%%%%%%%%%%%%%%%%%
% Discuss why filter-based approach is bad. The idea is it cannot
% filter enough data to gain speedup when majority of data is positive.
%%%%%%%%%%%%%%%%%%%%%%%%%%%%%%%%%%%%%%%%%%%%%%%%%%%%%%%%%%%%%%%%%%%%%%%
\PP{Limitation \rom{2} -- frequent events}. 
The filtering technique used in the \syspp system~\cite{pp} relies on 
data reduction by the filter to achieve speedup.
Let's assume the system is processing $N$ frames and that the fraction of frames 
that is filtered and discarded by the filter is $r$.
Let the costs of running the filter and running the object detector be 
$C_{f}$ and $C_{o}$ per frame, respectively.
To obtain a speedup, the data reduction rate must satisfy this constraint:
\[ N(C_{f} + (1 - r) \cdot C_{o}) < NC_{o} \equiv r > \frac{C_{f}}{C_{o}} \]
This constraint is not met by frequent events (\eg \qone in~\cref{mov:qone}).
In this case, since $r$ is small, the filter slows down the overall pipeline
since it adds additional overhead.
As a result, \syspp is slower than \sysnaive 
(\ie naively running object detector on every frame) for frequent queries like \qone.
~\syspp only provides a 0.93\X speedup compared to \sysnaive in this case.
Instead, for rare queries like \qthree, \syspp is able to provide a 1.44\X speedup
compared to \sysnaive.
We seek to dynamically adjust the query execution pipeline based on the 
estimated frequency of the event.

%%%%%%%%%%%%%%%%%%%%%%%%%%%%%%%%%%%%%%%%%%%%%%%%%%%%%%%%%%%%%%%%%%%%%%%
% Discuss why specialization model is bad in terms of accuracy. Idea is
% it mostly provides very bad accuracy.
%%%%%%%%%%%%%%%%%%%%%%%%%%%%%%%%%%%%%%%%%%%%%%%%%%%%%%%%%%%%%%%%%%%%%%%
\PP{Limitation \rom{3} -- difficult-to-detect objects}. 
\sysblaze~\cite{blazeit} uses a specialized model to directly return aggregates
(\eg number of cars in an image). 
This approach does not generalize to complex visual datasets.
The reasons are twofold.
First, the specialized model is designed to be shallow for fast execution.
So, it is unable to learn complex patterns.
Second, it relies on an ad-hoc subset of videos for training,
so the lack of positive examples greatly affects the quality of the model.

As shown in~\cref{tb:mov:blazeit_perf}, \sysblaze returns precise answers for
easy-to-answer queries.
However, it has a lower recall metric.
For hard-to-answer queries (\eg \qthree), the specialized model does not offer 
useful results. 
So, the system instead falls back to the object detection model.
In this case, \sysblaze runs the specialized model, resulting in lower
performance than \sysnaive.
In contrast, \sys is capable of selecting an optimal plan with good accuracy
and performance metrics.

\begin{table}[t]
  \renewcommand{\arraystretch}{1.2}
  \centering
  \begin{tabular}{@{}lccc@{}}

  \textbf{Difficulty}
  & \textbf{Precision}
  & \textbf{Recall} 
  & \textbf{Throughput} \\
  
  \toprule
  
  Easy
  & 97.72\%
  & \textbf{67.98\%}
  & 12.98\X \\
  
  Hard
  & 100.00\%
  & 100.00\% 
  & \textbf{0.93\X} \\
  
  \end{tabular}
  \caption{\textbf{\sysblaze vs \sysnaive} -- 
  Key metrics of \sysblaze with respect to a 
  \sysnaive system that only uses the heavyweight object detector.
  }
  \label{tb:mov:blazeit_perf}
\end{table}

\begin{figure}[t]
\begin{subfigure}[t]{0.49\linewidth}
  \centering
  \includegraphics[width=0.8\linewidth]{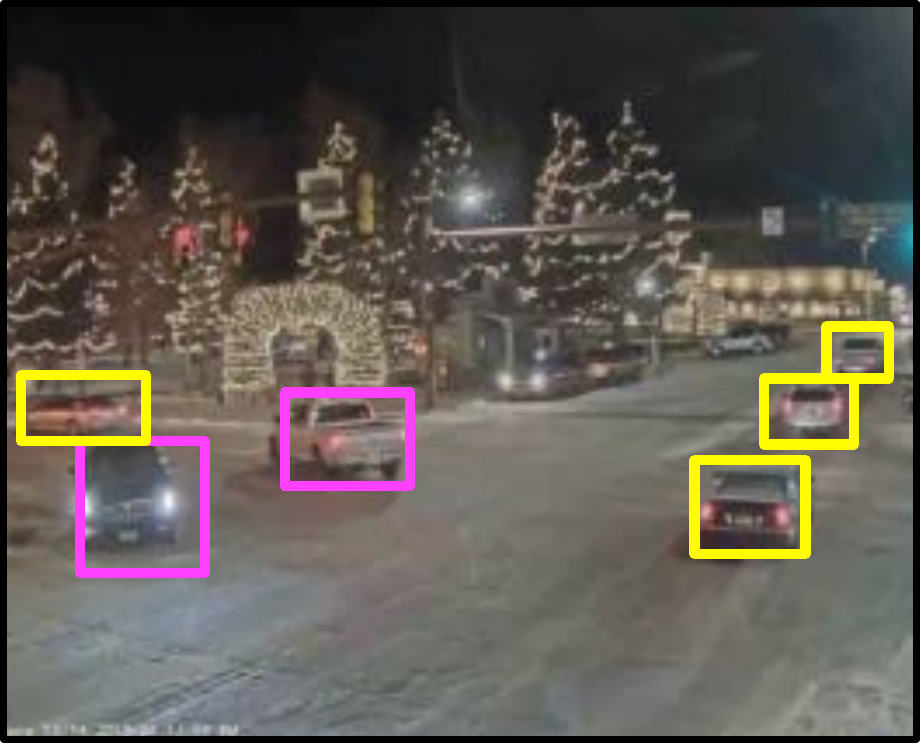}
  \caption{\footnotesize{Results of oracle \ep - 1\X speedup.}}
  \label{fig:mov:pred_image:oracle}
\end{subfigure}
\begin{subfigure}[t]{0.49\linewidth}
  \centering
  \includegraphics[width=0.8\linewidth]{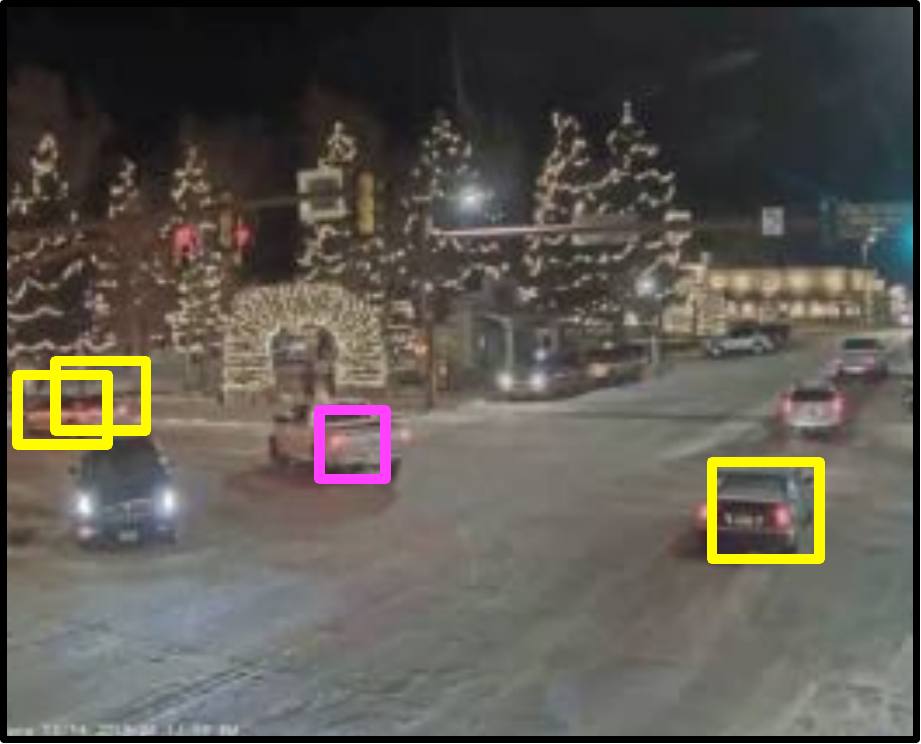}
  \caption{\footnotesize{Results of shallow \ep - 6\X speedup.}}
  \label{fig:mov:pred_image:approx}
\end{subfigure}
\caption{\textbf{Objects Detection Results} -- 
Objects detected by (a) oracle, and (b) shallow \ep.}
\label{fig:mov:pred_image}
\end{figure}

%%%%%%%%%%%%%%%%%%%%%%%%%%%%%%%%%%%%%%%%%%%%%%%%%%%%%%%%%%%%%%%%%%%%%%%
% The main idea here is approximate model is still very useful for some
% queries. 
%%%%%%%%%%%%%%%%%%%%%%%%%%%%%%%%%%%%%%%%%%%%%%%%%%%%%%%%%%%%%%%%%%%%%%%
%
\PP{Our Approach}. 
In~\cref{fig:mov:pred_image}, we show two prediction results from our \ei
technique (the oracle object detection \ep and a shallow \ep,
respectively).
We observe that the faster \ep is still able to capture the presence of cars, 
but it is less accurate in two ways.
First, the bounding boxes are not accurate, so multiple bounding boxes are 
returned for the same object.
Second, it tends to miss hard-to-detect objects (\eg objects far away or 
objects with lights).
If a user queries for an image with exactly four cars, \sys uses the oracle 
exit point to satisfy the precision requirement. 
However, if the user is only interested in images with cars, \sys uses the
faster exit point to obtain a 6\X speedup.
We design \sys so that it carefully chooses the optimal query execution plan 
for every \chunk of the video to deliver higher accuracy and speedup.

\begin{figure}[t]
  \centering
  \includegraphics[width=0.8\columnwidth]{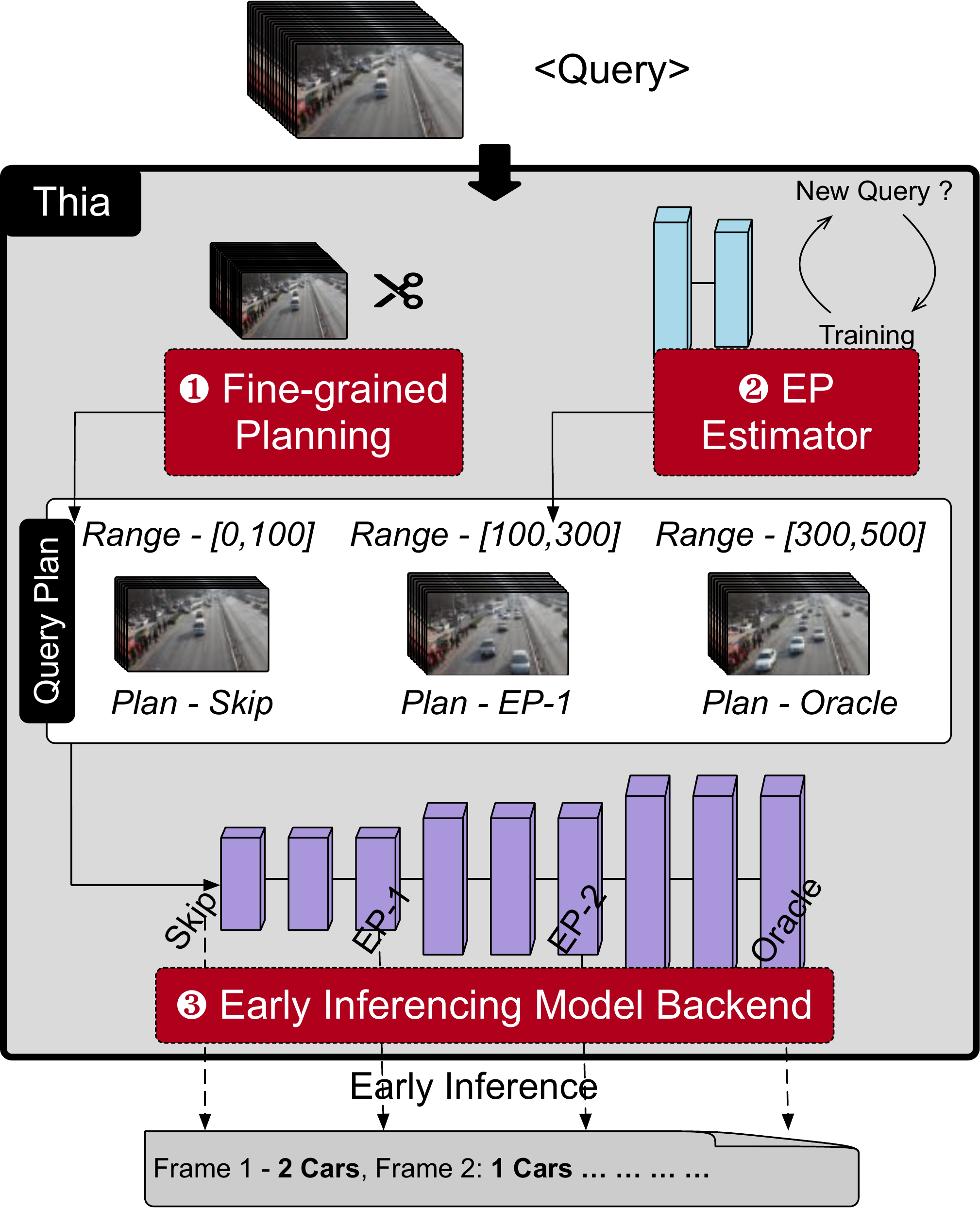}
  \caption{\textbf{System Overview} -- The two major components of \sys are:
  (1) \plan and (2) \exec.
  While the \plan relies on \ds and \me techniques, the \exec performs \ei.
  }
  \label{fig:sys}
\end{figure}

\section{System Overview}\label{sec:sys}
\cref{fig:sys} illustrates the architecture of \sys. 
%
%It consists of two components:
%(1) \plan, and
%(2) \exec.
%

%%%%%%%%%%%%%%%%%%%%%%%%%%%%%%%%%%%%%%%%%%%%%%%%%%%%%%%%%%%%%%%%%%%%%%%
% Discuss how dynamic sampling chooses a fine-grained plan.
%%%%%%%%%%%%%%%%%%%%%%%%%%%%%%%%%%%%%%%%%%%%%%%%%%%%%%%%%%%%%%%%%%%%%%%
\PP{\ding{182} \ds.}
When the system gets a query, the \plan uses the \ds technique to construct
a query execution plan.
It first splits the entire video into a set of small {\chunk}s.
The size of a \chunk is determined dynamically at runtime (covered
in~\autoref{sec:plan}).
For each \chunk, the \plan chooses the optimal plan (\ie when to stop inference
in the model).
Such a fine-grained query plan enables \sys to deliver higher accuracy and
throughput compared to a coarse-grained plan for the entire video.
A naive technique for picking the plan consists of running the model
on a set of sampled frames from the chunk.
While the fine-grained plan reduces the query \execover
(\sysei in~\cref{tb:back:sys_variants}), it increases the query \sampleover,
which hurts the overall query processing time (discussed
in~\autoref{sec:eval:over}). 
To reduce the \sampleover, \sys instead leverages a more lightweight 
\me technique.
%

%%%%%%%%%%%%%%%%%%%%%%%%%%%%%%%%%%%%%%%%%%%%%%%%%%%%%%%%%%%%%%%%%%%%%%%
% Explain the model estimation technique is to reduce the query
% planning cost. In addition to that, explain a little bit that it 
% requires ad-hoc training for different queries.
%%%%%%%%%%%%%%%%%%%%%%%%%%%%%%%%%%%%%%%%%%%%%%%%%%%%%%%%%%%%%%%%%%%%%%%
\PP{\ding{183} \me.}
\sys uses \me and \ds techniques in tandem to reduce the overhead of the \plan.
The \plan uses a shallow neural network to directly estimate when to 
short-circuit the inference in an \ei model.
It trains an \ep estimator for every unique query executed in the system.
We discuss how \sys obtains data for training the \me model
in~\autoref{sec:plan}.

%%%%%%%%%%%%%%%%%%%%%%%%%%%%%%%%%%%%%%%%%%%%%%%%%%%%%%%%%%%%%%%%%%%%%%%
% Go over the early inferencing backend.
%%%%%%%%%%%%%%%%%%%%%%%%%%%%%%%%%%%%%%%%%%%%%%%%%%%%%%%%%%%%%%%%%%%%%%%
\PP{\ding{184} \ei.}
The fine-grained query plan constructed by the \plan consists of a 
list of \chunk{s} and the model chosen for each \chunk.
For example, \sys may skip frames $0$ through $100$, run \ep-1 on frames
$101$ through $300$, and evaluate the oracle \ep (\ie \ep-3) on frames $300$
through $500$.
The \exec takes this query plan and uses the \ei technique to deliver different
accuracy-performance tradeoffs with a single model.

\section{\ei}\label{sec:ei}
In this section, we present the \ei technique. 
We first provide an overview of this technique in~\autoref{sec:ei:design}.
We then illustrate its utility using a case study with 
Faster-RCNN~\cite{faster-rcnn} in~\autoref{sec:ei:case}.

%%%%%%%%%%%%%%%%%%%%%%%%%%%%%%%%%%%%%%%%%%%%%%%%%%%%%%%%%%%%%%%%%%%%%%%
% Give a big overview of early inferencing technique.
%     1. Why it is necessary.
%     2. How do we enable it from a high-level.
%%%%%%%%%%%%%%%%%%%%%%%%%%%%%%%%%%%%%%%%%%%%%%%%%%%%%%%%%%%%%%%%%%%%%%%
\subsection{Overview}\label{sec:ei:design}

%%%%%%%%%%%%%%%%%%%%%%%%%%%%%%%%%%%%%%%%%%%%%%%%%%%%%%%%%%%%%%%%%%%%%%%
% Explain the goal of early inferencing technique (what it really wants
% to achieve. Compared to having multiple models, why this technique
% is still needed.
%%%%%%%%%%%%%%%%%%%%%%%%%%%%%%%%%%%%%%%%%%%%%%%%%%%%%%%%%%%%%%%%%%%%%%%
%
We seek to construct a \textit{single} model with multiple exit points
wherein the inference may be short-circuited to improve performance at the
expense of accuracy.
We do \textit{not} want to construct a collection of models to accomplish this
goal.
The \plan dynamically adjusts the \ep based on the query.
If the query is relatively easy to answer, \sys delivers higher speedup by
stopping the inference earlier (while returning accurate results).
We discuss how \sys estimates the correct \ep for a chunk
in~\autoref{sec:plan}.
In this section, we focus on how we construct a model with multiple \eps.

%%%%%%%%%%%%%%%%%%%%%%%%%%%%%%%%%%%%%%%%%%%%%%%%%%%%%%%%%%%%%%%%%%%%%%%
% Discuss the current object detection networks. Then explain our 
% intuition about how to apply early inferencing on those networks.
%%%%%%%%%%%%%%%%%%%%%%%%%%%%%%%%%%%%%%%%%%%%%%%%%%%%%%%%%%%%%%%%%%%%%%%
As discussed in~\autoref{sec:back:obj_det}, object detection models usually
rely on a backbone network that is based on a state-of-the-art image
classification model (\eg ResNet-50~\cite{resnet-50} and VGG-16~\cite{vgg-16}).
Since these classification models are tailored for high accuracy, 
they consist of a stack of compute-intensive layers that lead to lower 
inference throughput.
The layers in a backbone network are sequentially connected to each other.
Our key idea is to provide faster detection results with lower accuracy by  
using the features from earlier layers in the backbone network.
%

%%%%%%%%%%%%%%%%%%%%%%%%%%%%%%%%%%%%%%%%%%%%%%%%%%%%%%%%%%%%%%%%%%%%%%%
% Start with discussing some prior works of model cascading (I think
% this may add some credibility to this idea). By differentiating and 
% comparing to those works, I explain how our models are instrumented.
%%%%%%%%%%%%%%%%%%%%%%%%%%%%%%%%%%%%%%%%%%%%%%%%%%%%%%%%%%%%%%%%%%%%%%%
\PP{Model Cascading vs Early Inference:}
Researchers have proposed model cascades for face
recognition~\cite{face-cascade-1,face-cascade-2}.
Similar to \ei, in a model cascade, the features from earlier layers
in the backbone network are used for face recognition.
However, these techniques differ in two ways.
First, face recognition is a binary classification task (\ie face exists in
the image or not).
So, the additional classification layers are only instrumented in this approach.
%
% \AJ{We instead introduce exit points across the entire model ?}. 
%
Second, these efforts propose a bespoke architecture to construct the cascade.
We instead seek to support early inference in widely used object detection
models.

\begin{figure}[t]
  \centering
  \includegraphics[width=0.95\columnwidth]{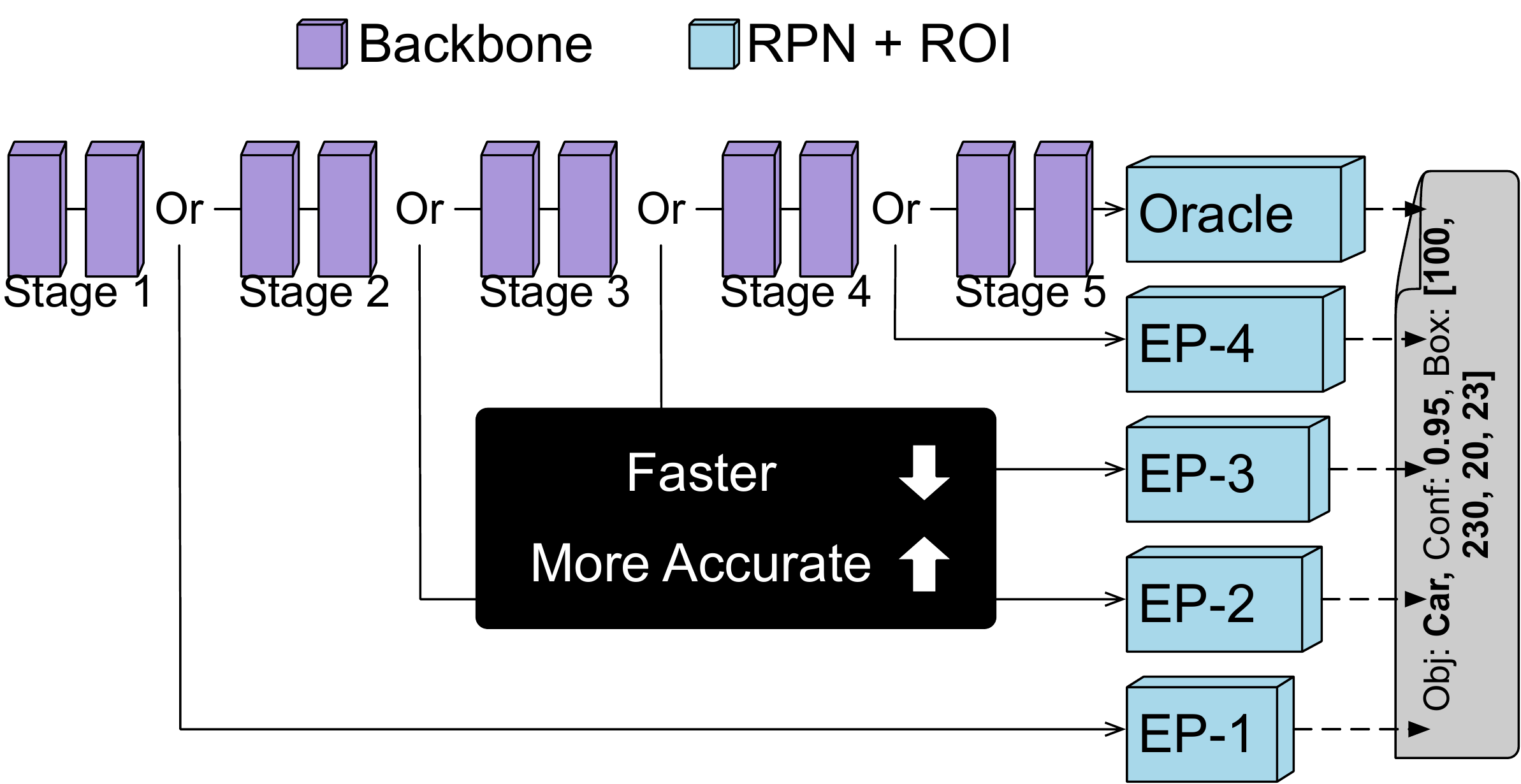}
  \caption{
  \textbf{\ei in Faster-RCNN} -- 
  Architecture of a Faster-RCNN model that supports early inference.}
  \label{fig:ei:case}
\end{figure}

To support \ei with a general-purpose object detection model,
we introduce additional \rpn and \roi
units in the earlier layers of the backbone network.
%\AJ{RPN and ROI -- Expand}
%
We modify the number of parameters in these units so that they operate on the
feature tensor emitted by the backbone network.
As shown in ~\cref{fig:ei:case}, for a given input, \plan may choose to
short-circuit the inference using the newly added units to speed up inference.

%%%%%%%%%%%%%%%%%%%%%%%%%%%%%%%%%%%%%%%%%%%%%%%%%%%%%%%%%%%%%%%%%%%%%%%
% Case study on Faster-RCNN. Give a concrete example.
%%%%%%%%%%%%%%%%%%%%%%%%%%%%%%%%%%%%%%%%%%%%%%%%%%%%%%%%%%%%%%%%%%%%%%%
\subsection{Case Study: Faster-RCNN}\label{sec:ei:case}
Faster-RCNN is a state-of-the-art object detector~\cite{faster-rcnn}.
We now discuss how we extend this model to support \ei.
We next describe how to generalize the training process to other models.

%%%%%%%%%%%%%%%%%%%%%%%%%%%%%%%%%%%%%%%%%%%%%%%%%%%%%%%%%%%%%%%%%%%%%%%
% Explain how we construct the model and a few design choices that 
% we have made.
%%%%%%%%%%%%%%%%%%%%%%%%%%%%%%%%%%%%%%%%%%%%%%%%%%%%%%%%%%%%%%%%%%%%%%%
\PP{Faster-RCNN with \ei}: 
The backbone network of Faster-RCNN is the ResNet-50~\cite{resnet-50} model.
%
%This \ei model is also used for all our evaluations in~\autoref{sec:eval}.
%
ResNet-50 consists of five stacked compute blocks,
so we extend this model to support five EPs (we could support fewer or
additional EPs if needed by instrumenting other layers of the backbone network). 
The default output of the model corresponds to the fifth EP.
We add four additional EPs that provide a wide set of throughput-accuracy
tradeoffs (\ie EP-1, EP-2, EP-3, and EP-4 in~\cref{fig:ei:case}).
We refer to EP-5 as the oracle (since it is the output of the original model).
We preserve the structure of RPN and ROI units as is the case of the oracle.
However, we modify the first layer in these units to work with the output
tensors of the early EPs that vary in size.
~\cref{tb:ei:case:layer} lists the layer configuration of each EP
\footnote{We found that upsampling the input channel size to $2048$ does not 
improve accuracy since the features from earlier EPs are coarse.}.

\begin{table}[t]
  
  \begin{threeparttable}
  
  \renewcommand{\arraystretch}{1.2}
  \centering
  \begin{tabular}{@{}lccccc@{}}

  \toprule

  \multirow{2}{*}{\textbf{\makecell{Exit Points}}}
  & \multicolumn{2}{c}{\textbf{Layer Config}}
  & \multicolumn{3}{c}{\textbf{Performance}} \\ 
  
  \cmidrule{2-3} \cmidrule{4-6} 
  
  & Channel
  & Kernel
  & Speedup
  & TPr
  & FNr \\
  
  \toprule
  
  \textbf{EP-1}
  & 64
  & 3 \X 3 
  & 6.90\X 
  & 87.99\% 
  & 42.70\% \\
  
  \textbf{EP-2}
  & 256
  & 3 \X 3 
  & 2.62\X 
  & 91.65\% 
  & 26.95\% \\
  
  \textbf{EP-3}
  & 512
  & 3 \X 3 
  & 2.46\X 
  & 95.52\% 
  & 16.22\% \\
  
  \textbf{EP-4}
  & 1024
  & 3 \X 3 
  & 1.97\X 
  & 98.17\%
  & 6.56\% \\
  
  \textbf{EP-5 (Oracle)}
  & 2048
  & 3 \X 3 
  & 1.00\X 
  & 100.00\% 
  & 0.00\% \\
  
  \bottomrule
  
  \end{tabular}
  
  \footnotesize{
  \begin{tablenotes}
    \item TPr: True positive ratio. FNr: False negative ratio.
  \end{tablenotes}
  }
  
  \end{threeparttable}
  
  \caption{\textbf{\ei model knobs} -- the layer configuration and performance of each 
  object detection knob in \ei model.
  }
  \label{tb:ei:case:layer}
\end{table}

%%%%%%%%%%%%%%%%%%%%%%%%%%%%%%%%%%%%%%%%%%%%%%%%%%%%%%%%%%%%%%%%%%%%%%%
% Discuss how is the model trained by introducing the loss function
% we use for warming up.
%%%%%%%%%%%%%%%%%%%%%%%%%%%%%%%%%%%%%%%%%%%%%%%%%%%%%%%%%%%%%%%%%%%%%%%
\PP{Top-down training}: 
We adopt a novel top-down training technique for constructing models that
support early inference.
We start the training process with the following multi-loss function:
\[ L(x; Y) = \frac{1}{|E|}\sum_{e \in E} L(\{y_{c}\}, \{y_{t}\}; Y) \]
$E$ represents the set of exit points (including the oracle).
$L$ denotes the object detection loss function used in
Faster-RCNN~\cite{faster-rcnn}.
This training step tunes all \eps.

$E$ represents all possible object detection \eps, including the oracle \ep.
We begin with the oracle \ep.
The reasons for doing this are twofold.
First, the oracle gets the features emitted by the last stage of the backbone
network, so training this \ep ensures that all layers converge to the optimal state.
Second, we seek to ensure that the oracle \ep in the \ei model delivers the same
accuracy as that of the original model.
After training the oracle \ep, we freeze all the layer parameters in the backbone
network.
This ensures that fine-tuning the shallow EPs later does not affect the 
previously tuned \eps.
We gradually fine-tune the RPN and ROI units starting from \ep-4 through
\ep-1.

%%%%%%%%%%%%%%%%%%%%%%%%%%%%%%%%%%%%%%%%%%%%%%%%%%%%%%%%%%%%%%%%%%%%%%%
% Discuss the speedup and accuracy. And then introduce slightly why
% the query planning is needed.
%%%%%%%%%%%%%%%%%%%%%%%%%%%%%%%%%%%%%%%%%%%%%%%%%%%%%%%%%%%%%%%%%%%%%%%
\PP{Throughput-Accuracy Tradeoffs}.
~\cref{tb:ei:case:layer} lists the speedup of shallow \eps with respect to the 
the oracle \ep.
For a given video frame, this \ei model offers up to a $6.9\times$ speedup when we
stop the inference at the first EP.
~\cref{tb:ei:case:layer} also summarizes the true positive and false negative 
percentage of each \ep on the training dataset with respect to the oracle \ep.
These metrics are averaged across all categories.
Shallower \eps return more false negatives and fail to return a few true
positives.  
In other words, they are more likely to not return a positive frame instead of
misclassifying a negative frame.
If the system were to use a shallow \ep for the entire video or sequence of
images, the impact on query accuracy would be significant.
Instead, it must use the oracle \ep on some difficult chunks of the video.
We cover this \ds technique in~\autoref{sec:plan}.
In~\autoref{sec:eval}, we demonstrate that \sys has a tolerable accuracy 
loss using both \ei and \ds techniques.

%%%%%%%%%%%%%%%%%%%%%%%%%%%%%%%%%%%%%%%%%%%%%%%%%%%%%%%%%%%%%%%%%%%%%%%
% Discuss the generalization of the idea to other networks.
%%%%%%%%%%%%%%%%%%%%%%%%%%%%%%%%%%%%%%%%%%%%%%%%%%%%%%%%%%%%%%%%%%%%%%%

\begin{figure}[t!]
  \includegraphics[width=0.45\columnwidth]{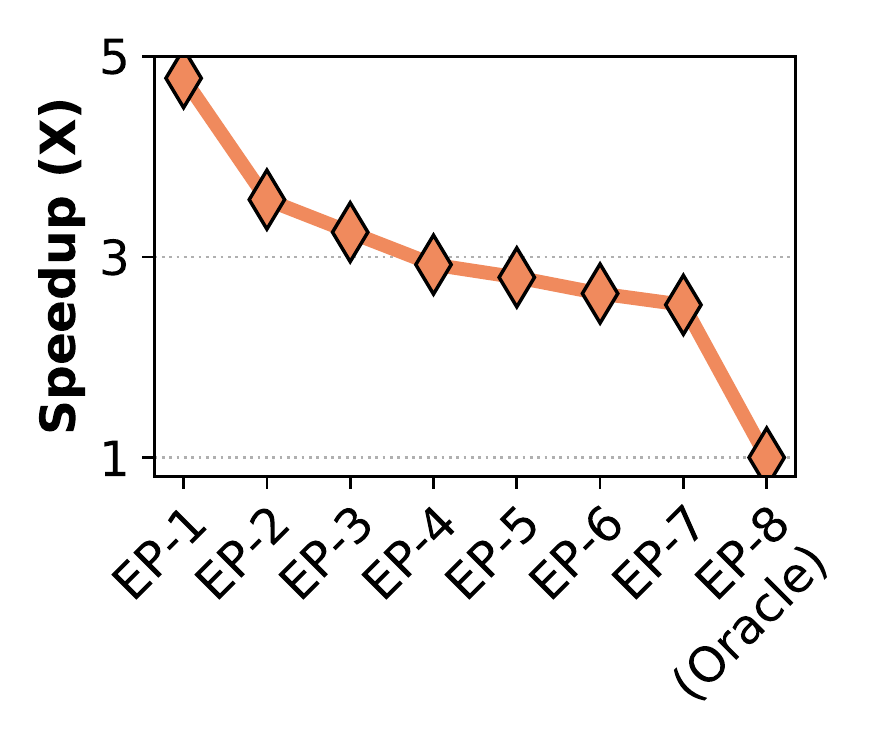}
  \caption{\textbf{Generalization of \ei} -- 
  Application of the \ei technique to a VGG-16 model for image classification.}
  \label{fig:ei:gen}
\end{figure}

\PP{Generalization.} 
The \ei technique generalizes to other models (\eg VGG-16~\cite{vgg-16}) and
other vision tasks (\eg image classification).
This is because most of these deep learning models contain similar backbone
networks that benefit from the \ei technique.
Furthermore, the number of \eps may be increased or decreased based on the
complexity of the model.

\cref{fig:ei:gen} illustrates another \ei model based on VGG-16 for an image
classification task that is trained on the Flower-102 dataset~\cite{flower-102}.  
Here, \ep-1 provides a $4.7$\X speedup compared to \ep-8 (\ie the oracle \ep).
By using all the eight \eps together, the system achieves a $2.7$\X speedup
compared to the oracle \ep with minimal accuracy loss.

\section{Query Planning}\label{sec:plan}

We present the \ds technique in this section.
In~\autoref{sec:plan:need}, we make the case for \ds.
In~\autoref{sec:plan:detail}, we discuss how \sys samples frames and constructs
\chunk{s} to apply this technique.
Lastly, in~\autoref{sec:plan:estimation}, we introduce the \me technique for 
reducing the \sampleover.

\begin{figure}[t]
  \centering
  \includegraphics[width=\linewidth]{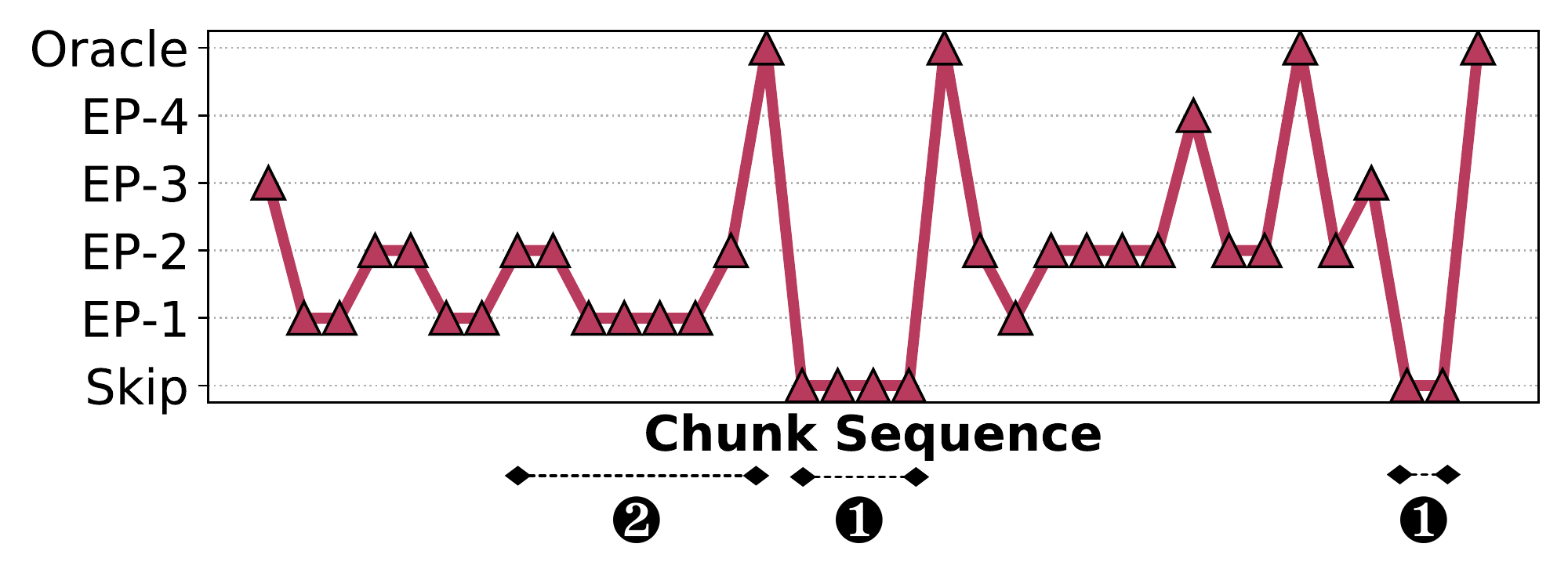}
  \caption{\textbf{Variation of Optimal \ep} -- The fastest, accurate \ep in an
  \ei model for a sequence of chunks in a video.}
  \label{fig:plan:plan_breakdown}
\end{figure}

\subsection{Motivation}\label{sec:plan:need}
As we discussed in~\autoref{sec:ei:case}, the \ei model contains a set of \eps.
The goal of the \plan is to choose an optimal (accurate and fast) \ep for every
fine-grained chunk of the video at runtime. 
Our key observation is that the optimal \ep changes at chunk granularity.
~\cref{fig:plan:plan_breakdown} illustrates the {\chunk}-level query plan for
\qfour in~\cref{mov:qtwo}.
The triangles in~\cref{fig:plan:plan_breakdown} represent the fastest (but
still accurate enough) \ep for every \chunk in the video.
This example shows that the optimal \ep constantly changes.
So, it is essential to dynamically adjust the query plan at \textit{runtime} 
to achieve both good accuracy and performance.

State-of-the-art systems (\eg \sysblaze~\cite{blazeit} and \syspp~\cite{pp})
take a coarse-grained approach to planning.
They choose a single plan for the entire video based on the accuracy of the model 
on a set of sampled frames.
The limitations of this technique are twofold.

\PP{Performance degradation}. 
Positive events tend to not appear in every chunk of the video 
(\ie selectivity of the predicate is typically high).
If we pick a static plan for the entire video, video chunks that are less 
likely to contain positive events or that contain easy-to-detect events are
passed to a more compute-intensive \ep.
Thus, the system does not leverage the opportunity to further improve performance
by either skipping those chunks or using less compute-intensive \eps for those
chunks.
With \ds, \sys uses a faster \ep or directly skips the entire
\chunk (\ding{182} in~\cref{fig:plan:plan_breakdown}).

\PP{Accuracy loss}. 
The distribution of the target event and the accuracy of the model vary across
the video.
A statically selected, shallow \ep will hurt accuracy by missing hard-to-detect
events.
As shown in~\cref{fig:plan:plan_breakdown}, some {\chunk}s require deeper \eps
to make accurate predictions (\ding{183} in~\cref{fig:plan:plan_breakdown}).
With \ds, \sys dynamically adjusts the plan based on the difficulty of 
detecting the target event.

\subsection{Chunking Algorithm}\label{sec:plan:detail}
When \sys gets a query, it first splits the given video into a set of {\chunk}s.
It then samples a set of frames from each \chunk and 
then evaluates the accuracy of all the \eps in the \ei model on these
sampled frames.
Using these results, the system selects the best \ep for each \chunk.
Lastly, it executes the query using the selected plan.
The key components of the algorithm that \sys uses for chunking videos are as follows: 

\begin{algorithm}[t]

\small

\SetKw{Is}{is}
\SetKw{And}{and}
\SetKw{Or}{or}

\SetKwInOut{Input}{Input}
\SetKwInOut{Output}{Output}

\SetKwArray{exitpointlist}{\textbf{EP-List}}

\SetKwData{precision}{\textbf{P}}
\SetKwData{recall}{\textbf{R}}
\SetKwData{video}{\textbf{V}}

\SetKwFunction{len}{Length}
\SetKwFunction{recursesample}{GetQueryPlan}
\SetKwFunction{estimateratio}{EstimateSamplingRate}
\SetKwFunction{predict}{Predict}
\SetKwFunction{pickbestep}{PickBestEP}

\SetKwProg{func}{Function}{}{}

\Input{\textit{\video} - Video data. \\
       \textit{\exitpointlist} - The list of \eps in the \ei model. \\
       \textit{\precision} - Precision constraint of the query. \\
       \textit{\recall} - Recall constraint of the query. \\}

\Output{Return a list of fine-grained plans. \\}
  
video\_length $\gets$ \len{\video}
  
\tcp{estimate initial sampling rate.}
sampling\_rate $\gets$ \estimateratio{\video} \label{algo:plan:estimate:bound}

\tcp{optimize the query plan.}
\Return \recursesample{\video, \exitpointlist, \precision, \recall, sampling\_rate} \label{algo:plan:search:top_level} \newline

\func{\pickbestep{V\_sub, \exitpointlist, \precision, \recall}} {
  \Output{Return the optimal \ep under \precision and \recall constraints, and
   the rate of positive frames in the sampled subset.}
}

\func{\recursesample{\video, \exitpointlist, \precision, \recall, sampling\_rate}} {
  \Output{A collection of fine-grained plans. \\}
  
  \tcp{divide into smaller chunks.}
  sampling\_span $\gets$ $\frac{1}{sampling\_rate}$ \\
  V\_sub $\gets$ \video[$0$, $1 \cdot sampling\_span$, $2 \cdot sampling\_span$ ...] \label{algo:plan:search:subset_const} \newline
  
  best\_ep, posi\_ratio $\gets$ \pickbestep{V\_sub, \exitpointlist, \precision, \recall} \label{algo:plan:search:pick} \newline
  
  \tcp{the collection of fine-grained plans.}
  plan $\gets$ \{\} 
  \If{(posi\_ratio \Is sufficient \And best\_ep \Is fastest) \Or (\len{\video}
  \Is small) \label{algo:plan:search:plan_fast}} { plan += \{ \video : best\_ep
  \} \\
  } \ElseIf {posi\_ratio \Is insufficient \label{algo:plan:search:discard}} {
    plan += \{ \video : skip \} \\
  } \Else {
    \For {V\_chunk $\in$ \video} {
      \tcp{double the sampling rate.}
      plan += \recursesample{V\_chunk, \exitpointlist, \precision, \recall, $2\cdot$sampling\_rate} \label{algo:plan:search:recurse}
    }
  }
  \Return plan
} 

\caption{Fine-grained query planning.}
\label{algo:plan:search}  
\end{algorithm}

\PP{\ding{182} Hierarchical Chunking}. 
The two key decisions made by the \plan are:
 (1) \chunk size, and 
 (2) sampling rate (\ie the number of frames to pick from a chunk).
The system delivers higher accuracy with a higher sampling rate since more
samples allow it to better estimate the optimal \ep for each chunk.
However, this hurts throughput since the system must evaluate the model's
behavior on more frames, thereby increasing \sampleover.
Choosing the \chunk size is also a challenging task.
This is because the duration of an event varies based on the video,
so \sys must dynamically adjust the \chunk size at runtime.

To tackle these challenges, \sys takes a hierarchical approach for picking the 
\chunk size and the sampling rate for each \chunk.
It initially uses a large \chunk size and a low sampling rate.
This allows the \plan to gain a rough understanding of the contents of the
video based on the inference results collected using the sampled frames.
Based on this knowledge, it recursively adjusts the \chunk size and sampling 
rate.

~\cref{algo:plan:search} presents the hierarchical, recursive technique used by
the \plan.
As shown in~\cref{algo:plan:search:plan_fast}, the recursive algorithm stops
when the \chunk size is smaller than a threshold or if the fastest \ep has been
chosen for a given \chunk that contains enough positive frames.
In~\cref{algo:plan:search:pick}, the \texttt{PickBestEP} function returns the
rate of positive frames in a chunk (\ie \texttt{posi\_ratio}) that is obtained
from the oracle \ep.  
It is important to ensure that the \chunk has sufficient positive frames,
since the calculated precision and recall metrics of the \eps do not 
generalize well without sufficient positive frames.
% \AJ{Why?}.s
%
These constraints bound the \sampleover.
As shown in~\cref{algo:plan:search:discard}, if there are very few positive
frames, then \sys skips the entire \chunk to reduce both \sampleover and \execover.
Lastly, it gradually reduces the \chunk size and increases the sampling rate,
as shown in~\cref{algo:plan:search:recurse}.
The intuition is that if the system is not able to select a plan based on its
coarse-grained understanding of the chunk, it must sample more frames from that
chunk in the next iteration.
By using a small \chunk size, the \plan is able to adjust the plans quickly 
to transient events.
 
\PP{\ding{183} Sampling Rate Bounds}. 
The \plan gradually increases the sampling rate to improve the quality of its
plan.
However, this increases the \sampleover and may hurt the overall throughput
obtained with the plan.
This is because the decrease in the query execution time is not sufficient to
justify the increase in the optimization time.
To overcome this limitation, the \plan uses the following constraint to bound
the initial sampling rate (\cref{algo:plan:estimate:bound}):
\[ sampling\_rate * 2^{\ceil{\log{\frac{|V|}{100}}}} \leq 0.1 \]
Here, we assume that \chunk{s} must contain at least 100 frames and we seek to
bound the final sampling rate to $0.1$ even in the worst-case setting.
The maximum depth of the recursive algorithm is
$\ceil{\log{\frac{|V|}{100}}}$ (sampling rate is doubled in each iteration).

\begin{figure}[t]
  \centering
  \includegraphics[width=0.8\linewidth]{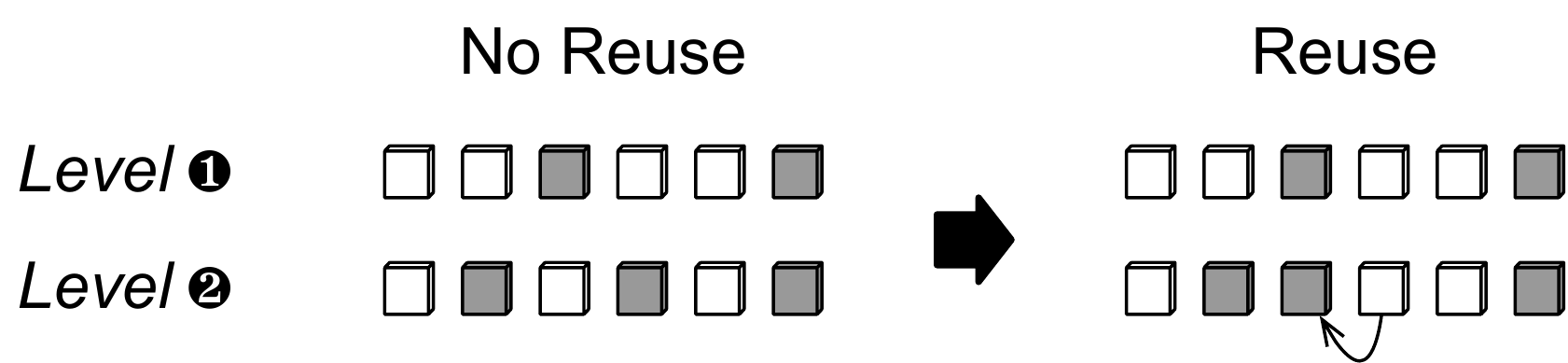}
  \caption{\textbf{Sampling results no reuse vs reuse} -- an illustrative example about
  sampling results about performance saving with no reuse and reuse.}
  \label{fig:plan:sampling_reuse}
\end{figure}

\PP{\ding{184} Memoization of Inference Results}.
In~\cref{algo:plan:search:subset_const}, the newly picked samples could be
different from those that have already been evaluated.
Evaluating all the \eps on a sample is expensive.
To reduce this overhead, the \plan memoizes the inference results and reuses
the results of \textit{nearby} frames.
This technique is illustrated in~\cref{fig:plan:sampling_reuse}.
Without memoization, the results for the second and fourth frames must be
obtained again when the sampling rate is increased in the next iteration.
\sys instead reuses the results of nearby frames within the same \chunk.
With memoization, it picks the cached results for the third frame instead of
running inference on the fourth frame.
Thus, it evaluates only the \eps on the second frame.

\PP{\ding{185} Evaluation-Based \ep selection}. 
To select the best \ep, as shown in~\cref{algo:plan:search:pick},
\sys evaluates all the \eps on the sampled frames and compares them
with the oracle \ep.
It picks the fastest \ep that provides $0.8$ precision and $0.8$
recall.
These constraints empirically offer maximal speedup with minimal accuracy loss
(\autoref{sec:eval:ei}).
Even with all of these optimizations, evaluating the \eps on a frame comes with
non-trivial \sampleover.
We next present the \me technique for further reducing the \sampleover.

\subsection{\me}
\label{sec:plan:estimation}
We seek to reduce the optimization time associated with the \ei technique.
As we present in~\autoref{sec:eval:over}, it is important to balance the
tradeoff between optimization time and execution time to improve the 
overall query processing time.

The \me technique consists of using a shallow, two-layer neural network instead
of the evaluation step in query planning.
The neural network directly returns the optimal \ep based on the backbone
features.
This allows the \plan to eliminate compute-intensive evaluation of all \eps.
For example, with Faster-RCNN, the inputs to the \me model are the features
emitted by the fifth stage of the \ei model.

To train this neural network, \plan uses $200$ images from the training dataset
of the \ei model along with the associated \ep decision.
For robust results, \sys must train a separate \me model for each query.
However, the overhead of training this model is tolerable because: 
(1) it is a one-time overhead for each query; and 
(2) training time for the \me model is negligible compared to total query
processing time due to the simple structure of the model.
We defer an empirical analysis of this optimization to~\autoref{sec:eval:me}.

\PP{Estimation-Based \ep selection}.
Since the \me model directly estimates the optimal \ep for a video frame,
it does not return precision and recall metrics for all the \eps.
So, the \plan extrapolates these metrics based on the estimated \ep.
We next discuss how this extrapolation is done.

Let us split the set of sampled frames into two subsets: 
(1) those that contain positive events as reported by the oracle \ep 
(\cref{algo:plan:search:plan_fast} in~\cref{algo:plan:search}), and
(2) those that do not contain positive events.
We define those two subsets as $S$ and $\widehat{S}$, respectively.
For a given video frame $x$, let us denote the \ep estimation model 
that outputs the optimal \ep for that frame by $OPT_{EP}(x)$.
The \plan estimates the number of true positives (TP),
false positives (FP), and false negatives (FN) for any \ep $k$ in the \ei as:
\begin{align*}
&TP_{k} = \sum_{x \for x \in S} g(x),
\qquad & g(x) = \begin{dcases}
  1, \text{\qquad if } k\geq OPT_{EP}(x)\\
  0, \text{\qquad otherwise}
\end{dcases}
\\
&FP_{k} = \sum_{x \for x \in \widehat{S}} g'(x),
\qquad & g'(x) = \begin{dcases}
  1, \text{\qquad if } k< OPT_{EP}(x)\\
  0, \text{\qquad otherwise}
\end{dcases}
\\
&FN_{k} = \sum_{x \for x \in S} g''(x),
\qquad & g''(x) = \begin{dcases}
  1, \text{\qquad if } k< OPT_{EP}(x)\\
  0, \text{\qquad otherwise}
\end{dcases}
\end{align*}
Our intuition is that a shallow \ep is less accurate than a deep \ep.
So, for a video frame $x$, the estimated optimal \ep (\ie $OPT_{EP}(x)$) returns 
correct results. 
Then, all \eps after the estimated optimal \ep (\ie k $\geq OPT_{EP}(x)$) 
should also return correct results, and vice versa.
Hence, in the case of positive events, a deeper \ep $k$ than the estimated optimal
\ep provides true positive prediction. 
On the other hand, a shallower \ep $k$ than the estimated optimal
\ep provides false negative prediction. 
In the case of negative events,  shallower \ep $k$ than the estimated optimal
\ep likely results a false positive.
With these projected metrics, the \plan derives the precision and recall
metrics for an \ep $k$ as:
\[Precision_{k} = \frac{TP_{k}}{TP_{k} + FP_{k}}, Recall_{k} = \frac{TP_{k}}{TP_{k} + FN_{k}}\]
Lastly, the \plan picks the fastest \ep that meets the precision and recall
constraints (\eg 0.8), as discussed earlier.

\section{Experimental Evaluation}\label{sec:eval}

We seek to answer the following questions in our evaluation:
\begin{itemize}
  \item How effective is the \ei technique in reducing the query
  processing time (\autoref{sec:eval:ei})?
  \item How much does each technique contribute to the overall
  performance (\autoref{sec:eval:factor})?
  \item How effective is the \ds compared to the coarse-grained planning
  (\autoref{sec:eval:plan})?
  \item What is the time spent on query planning and execution
  (\autoref{sec:eval:over})?  
  \item How effective is the \me technique in reducing the \sampleover
  (\autoref{sec:eval:me})?
  \item How effective is \sys compared to other state-of-the-art systems
  (\autoref{sec:eval:end2end})?
  \item How does the \ei technique compare against the model cascade technique
  (\autoref{sec:eval:mcopt})?
%  \item How close is the plan returned by the \plan to the optimal plan
%  (\autoref{sec:eval:perfect})?
\end{itemize}

\subsection{Experiment Setup}\label{sec:eval:setup}
\PP{Evaluated Systems}.
~\cref{tb:back:sys_variants} lists all the video analytics systems that we
compare in our analysis (including the variants of \sys).
In the \sysnaive system, we apply the oracle \ep on every frame.
We normalize the accuracy metrics of other systems against those of the
\sysnaive system.
We reimplement two other state-of-the-art systems in our framework for
comparative analysis: (1) \syspp~\cite{pp}, and (2) \sysblaze~\cite{blazeit}.
In our implementation, the \syspp system uses ResNet-34~\cite{resnet-50}
to filter out unrelated frames. 
The \sysblaze system uses a specialized model (ResNet-34) to accelerate queries.

To better understand the performance of \sys, we examine three variants of our
system:
(1) \syssingle uses only the \ds method with the oracle \ep.
(2) \sysmulti also uses the \ds method along with multiple \eps.
Specifically, we use Faster-RCNN models with three backbone networks: 
ResNet-18, ResNet-34, and ResNet-50~\cite{resnet-50} as three \eps. 
(3) \sysei is the closest variant of \sys. 
It uses the \ds along with the \ei technique (but does not use the \me
technique).

\PP{Datasets}.
We evaluate these systems on two datasets: 
(1) UA-DeTrac~\cite{ua_detrac}, and 
(2) Jackson-Town dataset from~\cite{blazeit}.
Both datasets are obtained from traffic surveillance cameras.
We focus on four vehicle categories in both datasets: 
\texttt{Car}, \texttt{Truck}, \texttt{Bus}, and \texttt{Others}.

\PP{Evaluation metrics}.
Similar to other video analytics systems~\cite{noscope, blazeit, pp, tahoma,
chameleon}, our evaluation normalizes the results with respect to the oracle model  
(Faster-RCNN model backed by ResNet-50).
So, we provide the \fonescore score calculated relative to the results of the
oracle model.
We also report separate precision and recall metrics for each query.
This is important since a user might require fine-grained accuracy requirements
(\eg low precision and high recall).
We assume that the decoded video is present on disk.

\PP{Queries}. 
To evaluate these systems, we use the four queries listed
in~\cref{tb:mov:queries}.
Based on the predicate, the frequency of true positive events and 
the difficulty of detecting those events vary.

\PP{Software and Hardware}. 
We implement \sys with the Detectron2~\cite{detectron2} framework in
PyTorch~\cite{pytorch}.
We evaluate these systems on a server with 44~CPU cores and 256~GB memory 
along with one Titan Xp GPU with 12~GB memory.

\PP{Model Training}. 
As discussed in~\autoref{sec:ei:case}, we construct the \ei model based on
Faster-RCNN.
We split the UA-DeTrac dataset into two parts: training and validation subsets. 
We train the \ei model on the the training subset.
We warm up the training process for $1$ epoch, and then each \ep in \ei model is
trained for $10$ epochs in a top-down manner.
Since the Jackson-Town dataset does not have ground-truth labels, we directly
apply the \ei model, which is tailored for the UA-DeTrac dataset.
For \sysmulti, we train three models: Faster-RCNN based on ResNet-18, 
Faster-RCNN based on ResNet-34, and  Faster-RCNN based on ResNet-50.
Each model is trained for 10 epochs.

To train the \me model, we use $1000$ images from the UA-DeTrac training
set.
We split these images into training ($200$ images) and validation subsets.
We construct the training so that the distribution of different \eps is
balanced.
The training data for this model consists of backbone features for those video
frames, and the output is the fastest \ep that is accurate enough.
We quickly train this shallow network for $20$ epochs. 

%%%%%%%%%%%%%%%%%%%%%%%%%%%%%%%%%%%%%%%%%%%%%%%%%%%%%%%%%%%%%%%%%%%%%%%
% EI + DS
%%%%%%%%%%%%%%%%%%%%%%%%%%%%%%%%%%%%%%%%%%%%%%%%%%%%%%%%%%%%%%%%%%%%%%%
\begin{figure}[t]
  \centering
  \includegraphics[width=0.7\linewidth]{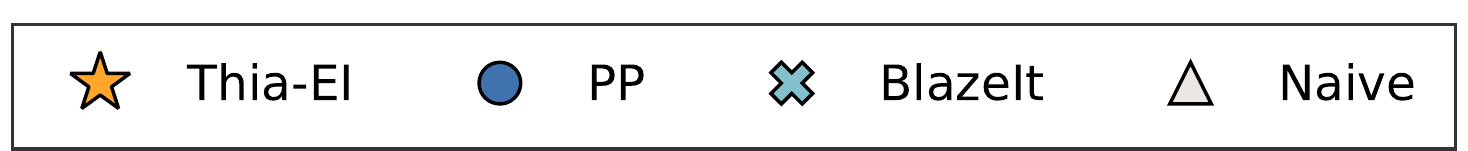}
  \includegraphics[width=0.9\linewidth]{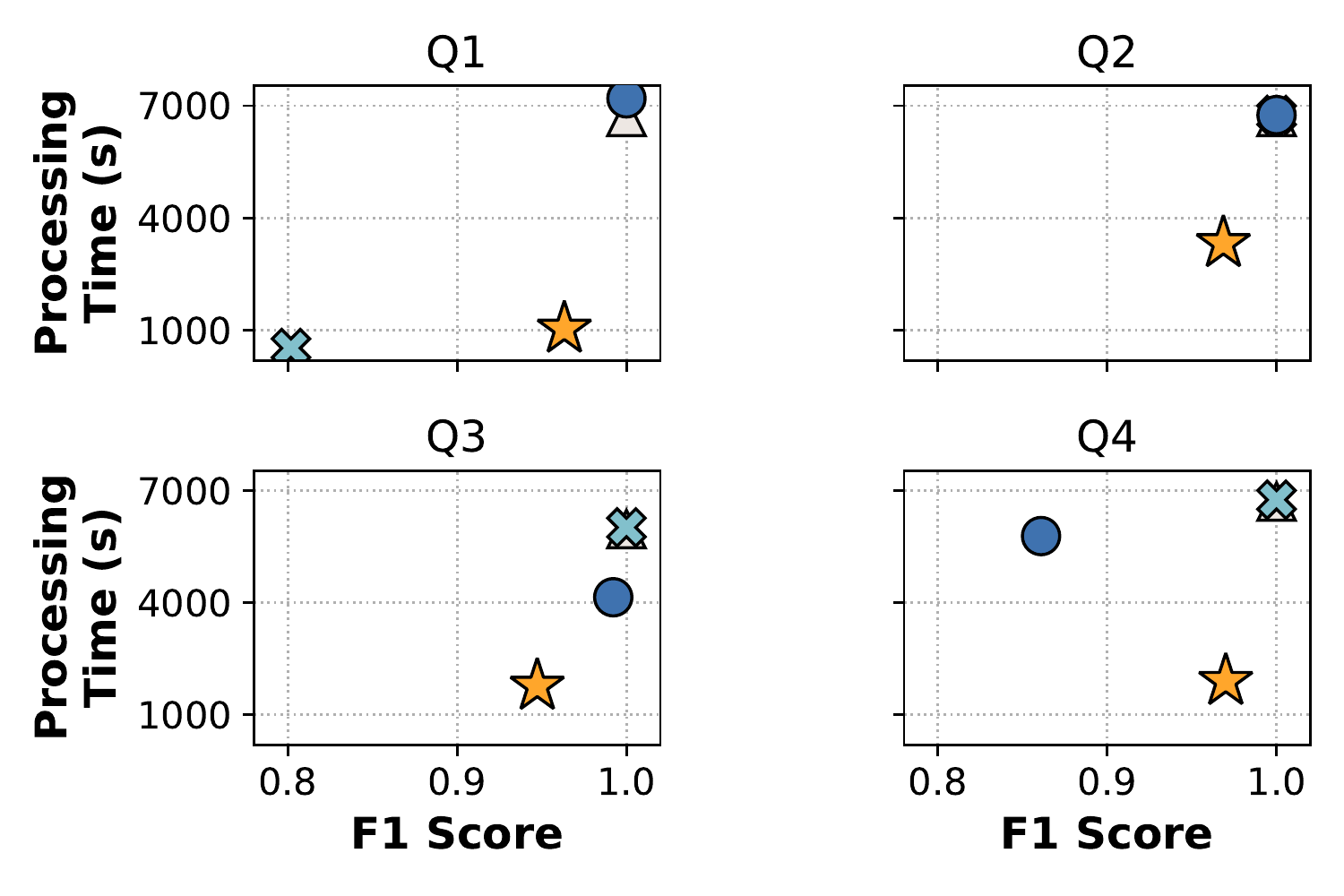}
  \caption{\textbf{Impact of \ei} --
  Query processing time and \fonescore scores delivered by \sysei, \syspp,
  and \sysblaze (bottom right corner represents the ideal system).}
  \label{fig:eval:ei:time_vs_f1}
\end{figure}

\subsection{Impact of \ei}\label{sec:eval:ei}
In this experiment, we compare the query processing time of \sysei to that of 
other video analytics systems.
The results are shown in~\cref{fig:eval:ei:time_vs_f1}. 
The bottom right corner represents the ideal case (faster execution with 
accurate predictions).

\PP{\sysei}.
The most notable observation is that \sysei outperforms other systems on most
queries. 
\sysei uses both \ei and \ds.
On \qone and \qtwo, since the fraction of frames filtered out is limited, 
using an extra specialized model before the object detector adds additional
execution overhead.
\sysei consistently reduces the total runtime and also delivers a higher
\fonescore score compared to other systems.
In particular, it is 2 -- 6\X faster than \sysnaive with a tolerable drop in
\fonescore score.

\PP{\sysblaze}.
On \qone, \sysblaze outperforms other systems with respect to query processing
time.
However, as we discussed in~\autoref{sec:mov}, its specialized model delivers
a lower \fonescore score.
On other queries, since the \fonescore score of the specialized model is too 
low to be useful, \sysblaze falls back to the oracle model.
Even though the specialized model is not effective, \sysblaze still evaluates
the query with the specialized model,
so the processing time of \sysblaze is higher than that of \sysnaive for \qtwo,
\qthree, and \qfour.

\PP{\syspp}.
\syspp reduces the processing time on \qthree and \qfour.
This is because these two queries focus on relatively rare events.
So, the model in \syspp is able to filter out a significant fraction of frames
to accelerate query processing. 
%

%
%In the following~\autoref{sec:eval:factor}, we discuss the detailed
% performance benefit breakdown from each component.

%%%%%%%%%%%%%%%%%%%%%%%%%%%%%%%%%%%%%%%%%%%%%%%%%%%%%%%%%%%%%%%%%%%%%%%
% Performance breakdown
%%%%%%%%%%%%%%%%%%%%%%%%%%%%%%%%%%%%%%%%%%%%%%%%%%%%%%%%%%%%%%%%%%%%%%%
\begin{figure}[t]
  \centering
  \includegraphics[width=\linewidth]{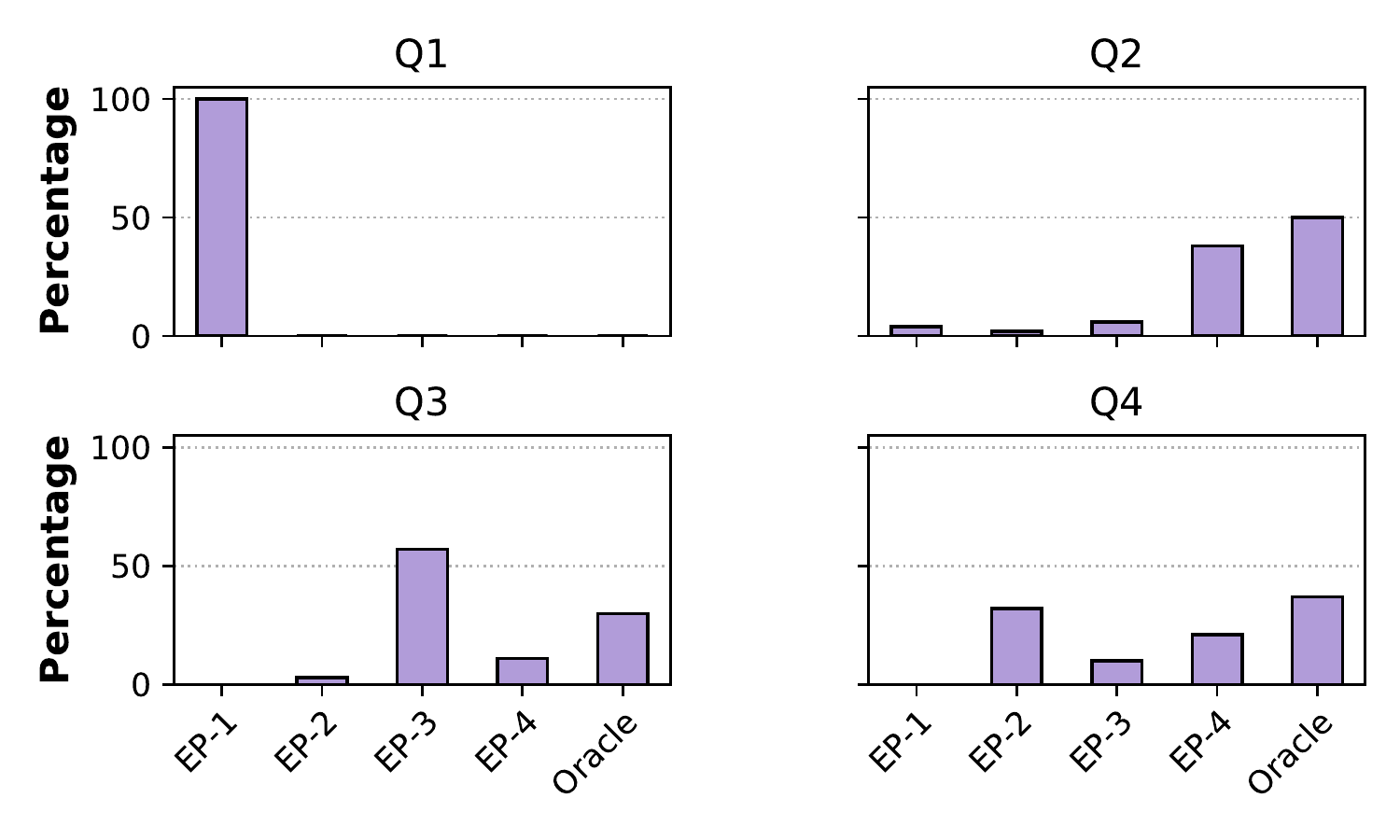}
  \caption{\textbf{Usage of \eps in \ei model} -- 
  Percentage of frames processed using the \eps in the \ei model.}
  \label{fig:eval:ei:ep_dist}
\end{figure}

\begin{figure}[t]
  \centering
  \includegraphics[width=\linewidth]{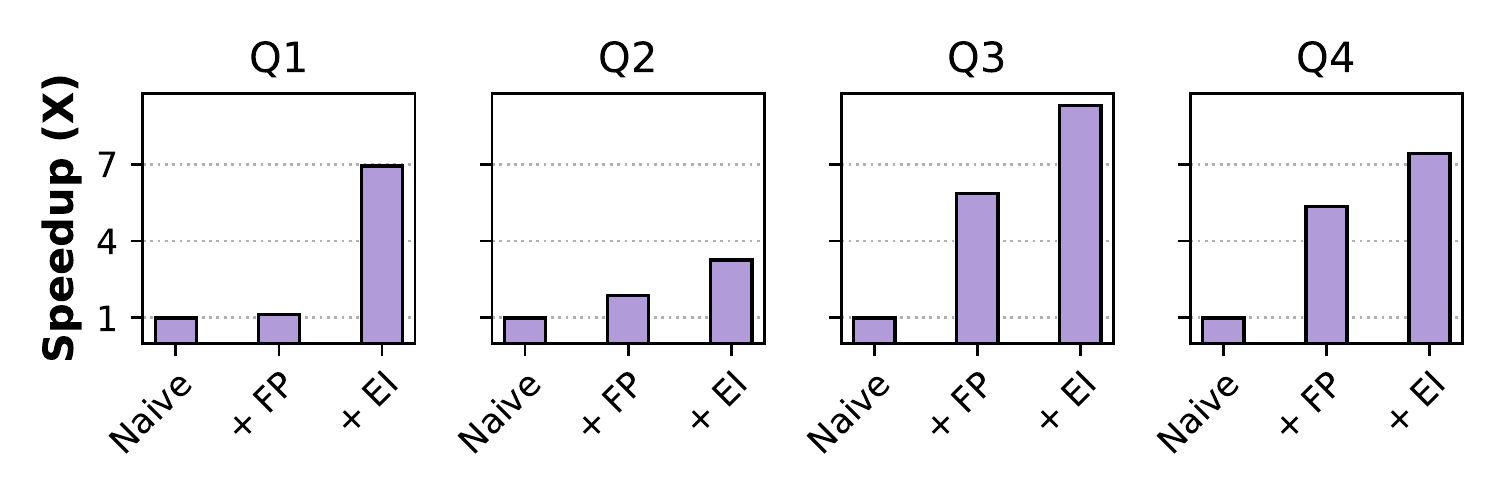}
  \caption{\textbf{Ablation study} -- Contribution of \ds and \ei techniques to
  the performance of \sysei.}
  \label{fig:eval:ei:factor_analysis}
\end{figure}

\subsection{Ablation Study}\label{sec:eval:factor}
We next examine how the \ep{s} in a model are used by \sysei while processing
queries.
The results of this experiment are shown in~\cref{fig:eval:ei:ep_dist}.
To better understand the contribution of each technique to the performance of
\sysei, we also conduct an ablation study.
We measure the \execover of \syssingle and \sysei to illustrate the benefits of
\ds and \ei techniques, respectively.
The results of this study are illustrated in~\cref{fig:eval:ei:factor_analysis}.

These two experiments demonstrate that: 
(1) \ds is able to adaptively choose the appropriate \ep for each chunk based on
the query and video frames, and 
(2) \ds and \ei techniques have significant impact for rare events; but, in
the case of frequent events, the speedup mainly comes from \ei.

On \qone, since positive events appear in majority of the video frames, the
impact of \ds is minimal.
When we add in the \ei technique, \sysei delivers higher speedup by using
shallow \eps for easy-to-detect events, as shown in~\cref{fig:eval:ei:ep_dist}.
\qone demonstrates an extreme scenario wherein the first \ep (EP-1) provides
correct predictions on all video frames.
In contrast, in the case of \qtwo, \sysei must use multiple \eps due to
harder-to-detect events.
Here, the \plan reduces the \execover by carefully choosing the \eps to use.
As shown in~\cref{fig:eval:ei:ep_dist}, while some frames are assigned to the
oracle \ep, other frames are assigned to shallow \eps to reduce \execover.
By reducing the \execover, \sysei delivers a 4\X speedup over \sysnaive.

Unlike queries focusing on frequent events, the \ds technique provides more
performance benefits in the case of queries related to rare events, 
because the \plan decides to skip some chunks during execution
(\cref{algo:plan:search:discard}).
On \qthree and \qfour, as shown in~\cref{fig:eval:ei:factor_analysis}, 
\ds leads to a 5\X speedup.
Nevertheless, the \ei technique is still useful for these queries.
\sysei carefully assigns certain video frames to shallow \eps to improve the
performance without losing accuracy.
The system is thus accelerated further by 3\X when the \ei technique enabled.

\begin{figure}[t]
  \centering
  \includegraphics[width=0.6\linewidth]{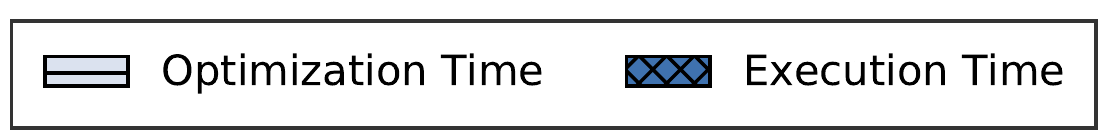}
  \includegraphics[width=0.9\linewidth]{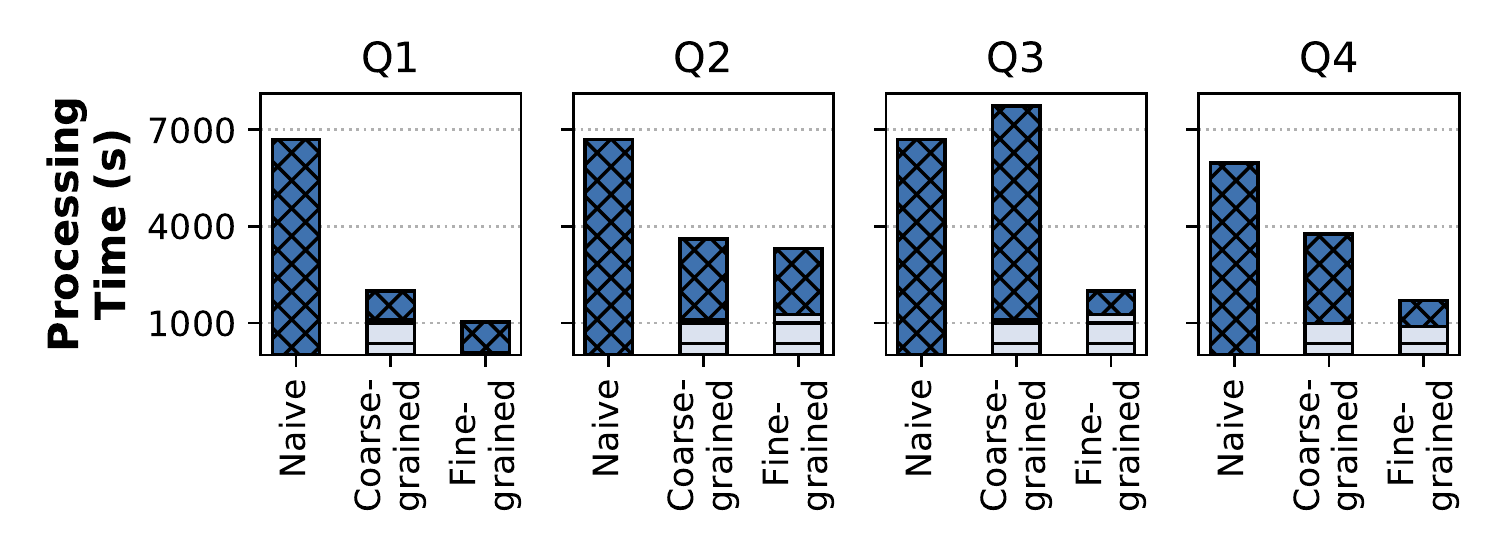}
  \caption{\textbf{Impact of \ds} --
  Breakdown of query processing time with fine-grained and coarse-grained
  planning techniques.}
  \label{fig:eval:plan:breakdown}
\end{figure}

\begin{figure}[t]
  \centering
  \includegraphics[width=0.8\linewidth]{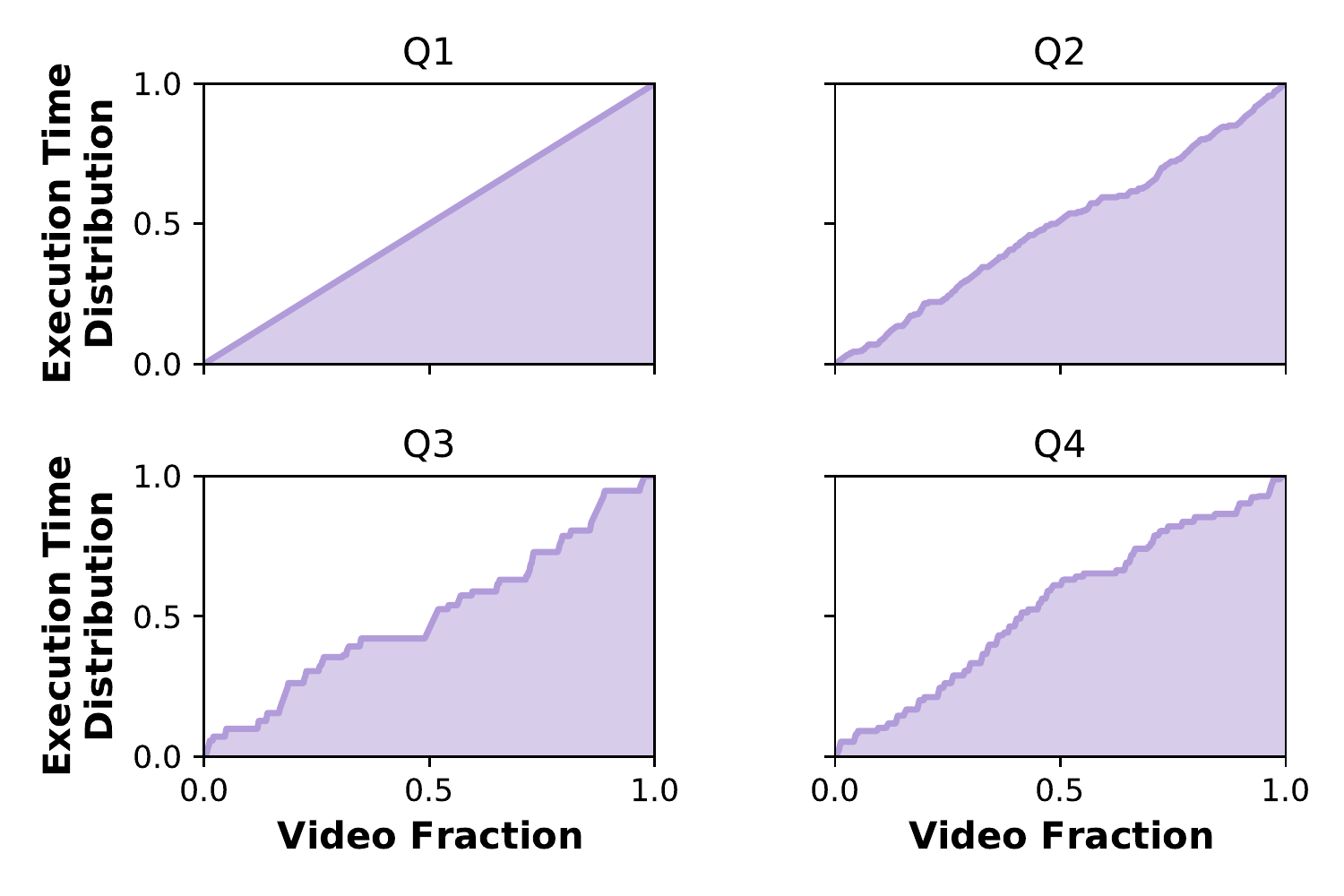}
  \caption{\textbf{Variation of Execution Time } --
  Variation of query execution time across the chunks in the video.}
  \label{fig:eval:plan:exec_dist}
\end{figure}

\subsection{Impact of \ds}~\label{sec:eval:plan}
We demonstrate the benefits of \ds by comparing \sysei against 
a system that uses coarse-grained planning with the same \ei model.
With coarse-grained planning, we evaluate all \eps on $10\%$ of the sampled
video frames and pick the \ep that meets the precision and recall
constraints.
We show a breakdown of the query processing time
in~\cref{fig:eval:plan:breakdown}.
On all queries, \ds provides a better query plans than coarse-grained planning so
that the execution time is consistently lower. 
Moreover, with optimizations like sampling rate bounds and memoization in \ds, it does not
incur higher optimization time than the naive coarse-grained planning approach.

We next measure the distribution of query execution time over the fraction
of the video being analysed in~\cref{fig:eval:plan:exec_dist}.
An even distribution (\eg \qone and \qtwo) suggests that the same query plan is
used for a large \chunk.
In contrast, an uneven distribution (\eg \qthree and \qfour) suggests
that the plan changes frequently across the video.

%%%%%%%%%%%%%%%%%%%%%%%%%%%%%%%%%%%%%%%%%%%%%%%%%%%%%%%%%%%%%%%%%%%%%%%
% System execution overhead breakdown
%%%%%%%%%%%%%%%%%%%%%%%%%%%%%%%%%%%%%%%%%%%%%%%%%%%%%%%%%%%%%%%%%%%%%%%
\begin{figure}[t]
  \centering
  \includegraphics[width=0.6\linewidth]{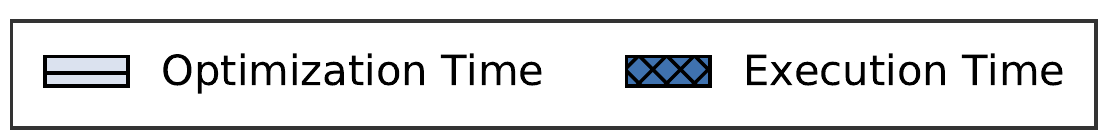}
  \includegraphics[width=\linewidth]{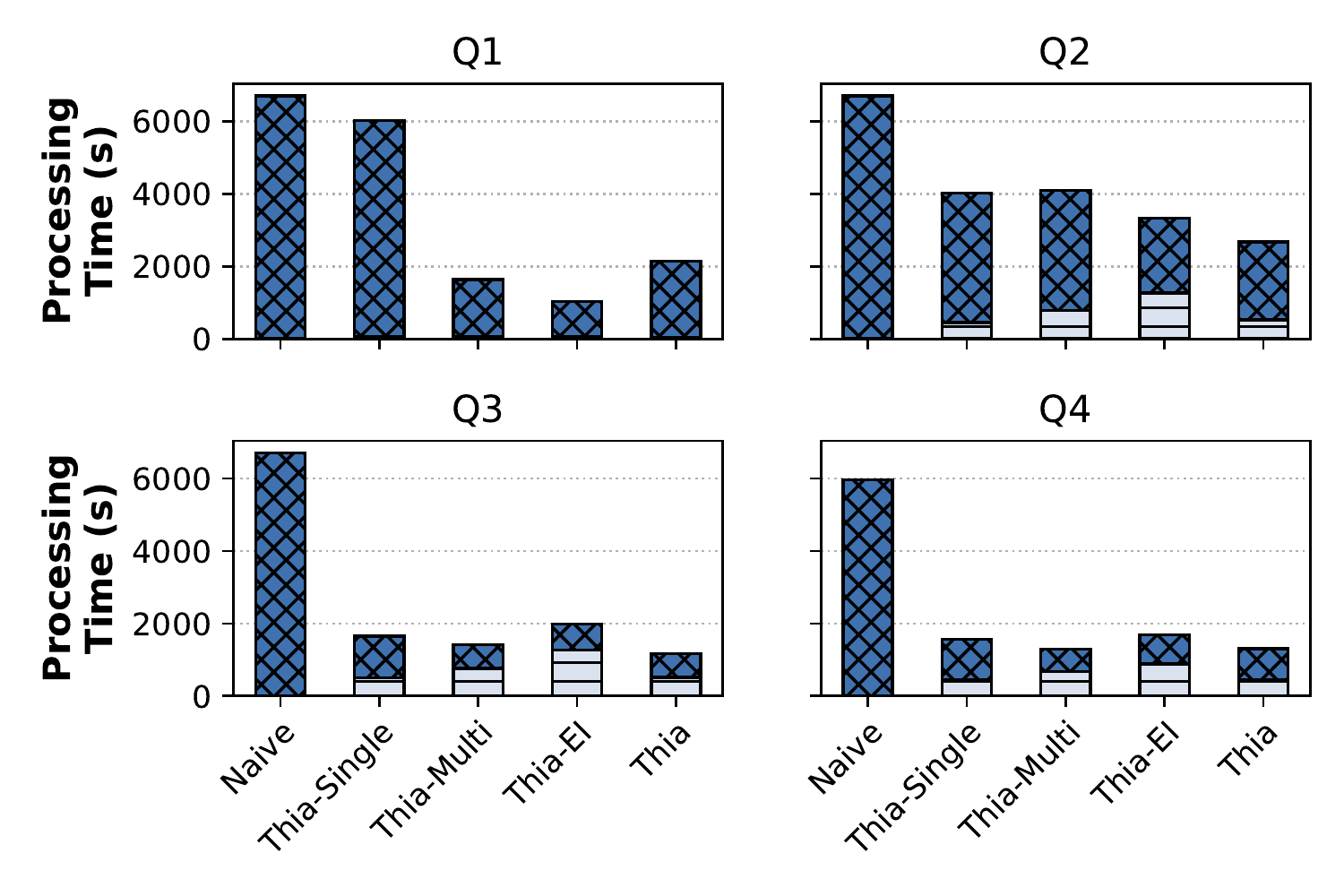}
  \caption{\textbf{Breakdown of query processing time} -- 
  Components of query processing time (\sampleover and \execover)
  associated with \sysnaive, \syssingle, \sysmulti, \sysei, and \sys.}
  \label{fig:eval:ei:sys_variant_breakdown}
\end{figure}

\subsection{Processing Time Breakdown}\label{sec:eval:over}
We now provide a breakdown of the processing time of \sysei and its
variants and compare it against \sysnaive.
Recall that all systems except for \sysnaive use the \ds technique.
Though \syssingle is only able to use the oracle \ep, it is able to skip frames
with no relevant events using \ds.

The results are shown in~\cref{fig:eval:ei:sys_variant_breakdown}.
Access to a set of \eps allows both \sysei and \sysmulti to reduce 
\execover in comparison to \syssingle and \sysnaive.
The reduction in \execover is more prominent compared to \syssingle 
for queries focusing on more frequent events.
While \sysmulti supports multiple \eps similar to \sysei,  
\sysei supports multiple \eps in a single model.
So, it has a lower GPU memory footprint, as shown in~\cref{fig:eval:mem}.
In addition to that, \sysei offers more flexibility in terms of creating and
selecting different \eps.
On \qtwo, due to the limited flexibility of \sysmulti, it has lower \execover and
also lower accuracy than \sysei.

The cons of using multiple \eps with \ds is the increase in \sampleover.
This is because the \plan has to evaluate all \eps to choose an optimal \ep.
As illustrated in~\cref{fig:eval:ei:sys_variant_breakdown}, the \sampleover of
\sysei and \sysmulti is consistently higher than that of \syssingle.
Increasing this flexibility (\ie adding more \eps) leads to higher sampling
overhead.
Thus, \sysei has higher \sampleover than \sysmulti and both have higher
\sampleover than \syssingle.
This motivates the need for reducing the \sampleover.
% by lowering the sampling overhead.
%
%\sysei relies on the \me technique to directly estimate an appropriate \ep 
%instead of running inference on the frames.
%
%We examine the efficacy of this technique in~\autoref{sec:eval:me}.

%%%%%%%%%%%%%%%%%%%%%%%%%%%%%%%%%%%%%%%%%%%%%%%%%%%%%%%%%%%%%%%%%%%%%%%
% Model estimation study
%%%%%%%%%%%%%%%%%%%%%%%%%%%%%%%%%%%%%%%%%%%%%%%%%%%%%%%%%%%%%%%%%%%%%%%
\begin{figure}[t]
  \centering
  \includegraphics[width=\linewidth]{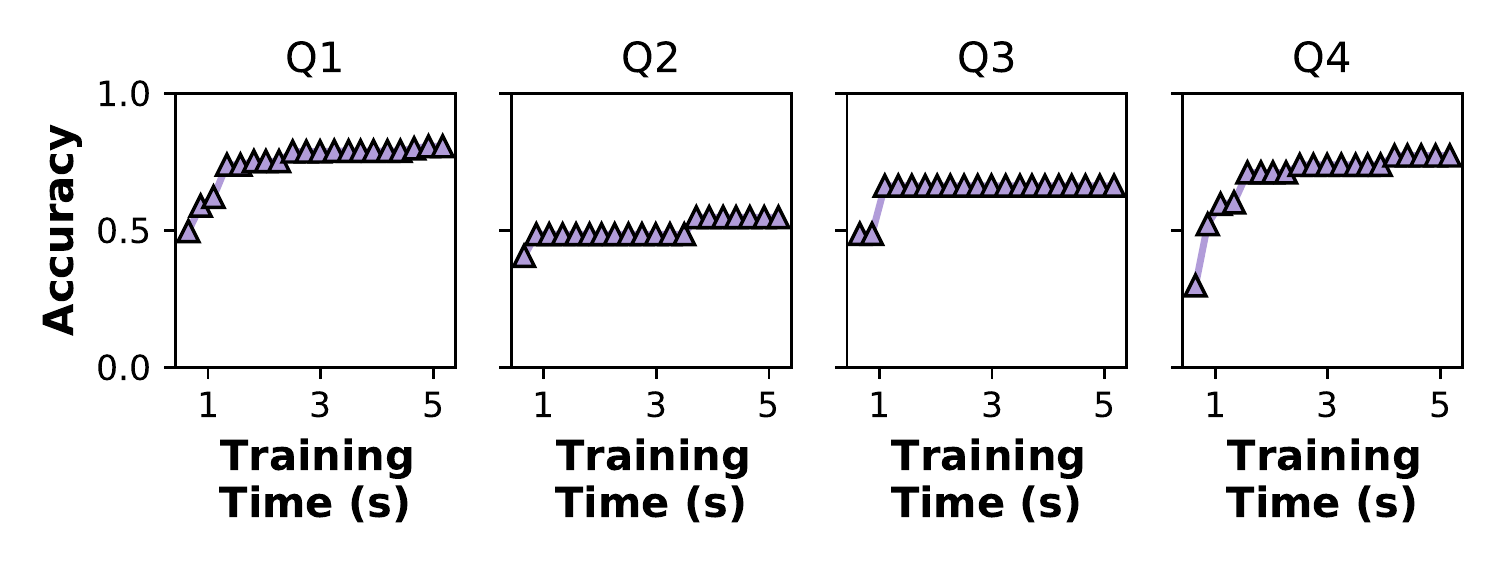}
  \caption{\textbf{Accuracy of the \me model} -- Variation in validation
  accuracy over training time.}
  \label{fig:eval:me:training}
\end{figure}

\begin{table}[t]
  \renewcommand{\arraystretch}{1.1}
  \centering
  \begin{tabular}{@{}crr@{}}

  \toprule
  
  \textbf{Query}
  & \multicolumn{1}{c}{\textbf{Under (\%)}}
  & \multicolumn{1}{c}{\textbf{Over (\%)}}\\
  
  \midrule
  
  \textbf{\qone}
  & 0.00
  & 27.29 \\
  
  \textbf{\qtwo}
  & 20.29
  & 21.42 \\
  
  \textbf{\qthree}
  & 28.96
  & 9.64 \\
  
  \textbf{\qfour}
  & 7.20
  & 14.40 \\
  
  \bottomrule
  
  \end{tabular}
  \caption{\textbf{Impact of \me technique} -- Accuracy of
  of \me technique relative to \sysei.}
  \label{tb:eval:me_acc}
\end{table}

\begin{figure*}
\includegraphics[width=0.45\textwidth]{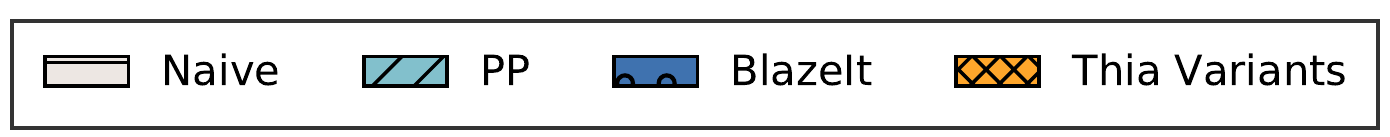}
\includegraphics[width=0.8\textwidth]{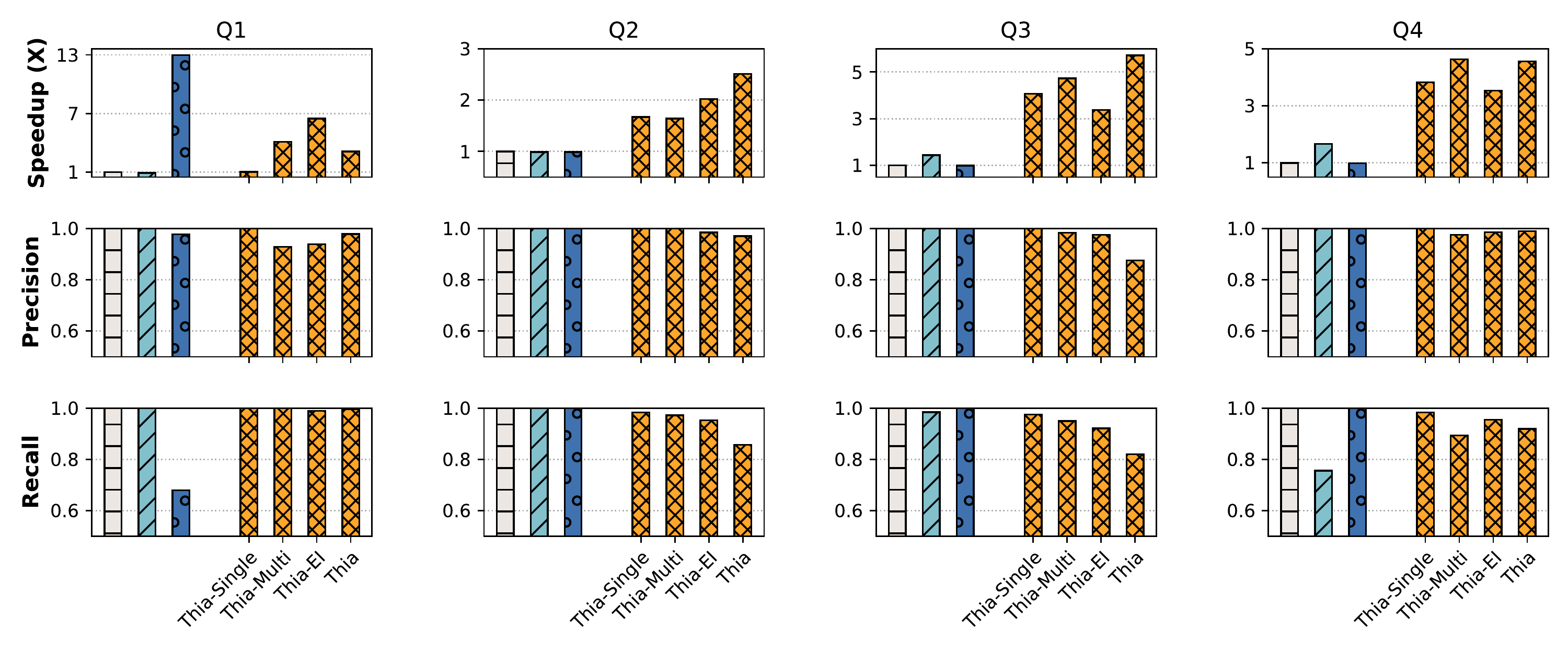}
\caption{\textbf{End-to-end Comparison} -- Comparative analysis of speedup,
precision, and recall metrics against state-of-the-art video analytics systems.}
\label{fig:eval:end2end}
\end{figure*}

\subsection{Impact of \me}\label{sec:eval:me}
\PP{Training Time}.
Since the \plan needs to train an estimation model for every unique query, 
we first quantify the training overhead of the \me technique.
~\cref{fig:eval:me:training} shows the variation in validation accuracy over
training time.
The model quickly converges since the \sys uses a small training set (200
samples) and a two-layer neural network for \me.
It takes less than ~$5$ seconds ($0.1$\% of total processing time) to train each
of these models for all queries.

\PP{\sampleover}.
We next investigate the efficacy of the \me technique in reducing the \sampleover.
We integrate the \me technique into \sysei to construct  \sys.
Using the \me technique, \sys is able to directly predict an appropriate \ep to
use for a \chunk.
In contrast, \sysei runs inference using all \eps during optimization phase to
select the \ep.
As shown in~\cref{fig:eval:ei:sys_variant_breakdown}, 
\sys cuts the \sampleover in half.
Recall that \sys uses the object detection \ep{s} for \ds.
So, the \me technique uses the backbone features for choosing a plan for
each \chunk.
We also measure the overhead of using \me.
This technique introduces a minimal additional overhead (18~s) even 
under the highest sampling rate
(processing time is in order of thousands of seconds).

\PP{\execover}.
Lastly, we discuss the impact of the \me technique on planning accuracy (\ie
choosing the optimal \ep) and \execover.
We measure the planning accuracy relative to \sysei.
In~\cref{tb:eval:me_acc}, \texttt{Under} and \texttt{Over} 
represent the percentage of chunks for which the \me technique returns a
shallower \ep and a deeper \ep than that returned by \sysei.
While shallower estimates hurt query accuracy, deeper estimates increase
\execover.
As shown in~\cref{fig:eval:ei:sys_variant_breakdown}, \execover increases only
negligibly for all queries except for \qone. 
Since the \me technique reduces \sampleover, the total processing time of \sys 
is lower than that of \sysei on all queries except for \qone.
This is because \qone can be accurately answered using the first \ep 
(\cref{fig:eval:ei:ep_dist}),
so deeper estimates increase \execover.
We discuss the impact on query accuracy in \autoref{sec:eval:end2end}.

%%%%%%%%%%%%%%%%%%%%%%%%%%%%%%%%%%%%%%%%%%%%%%%%%%%%%%%%%%%%%%%%%%%%%%%
% End to end comprehensive study
%%%%%%%%%%%%%%%%%%%%%%%%%%%%%%%%%%%%%%%%%%%%%%%%%%%%%%%%%%%%%%%%%%%%%%%

\subsection{End-to-End Comparison}\label{sec:eval:end2end}
We report the speedup, precision, and recall metrics with
respect to other state-of-the-art systems in~\cref{fig:eval:end2end}.
The bars on the left side represent three systems: 
(1) \sysnaive, (2) \sysblaze, and (3) \syspp.
The latter two video analytics systems use specialized models.
The bars on the right side represent variants of \sys that use one or more
techniques presented in this paper.

The most notable observation is that systems that have access to a set of \eps
deliver higher performance than those that have access to a single \ep.
Unlike \sysmulti, which maintains a collection of separate models, 
the \ei technique offers more flexibility in choosing the optimal \ep.
However, this technique increases the \sampleover.
We overcome this limitation using the \me technique.

\sys consistently delivers higher speedup than other systems.
Due to the inaccuracy of the \me technique, \sys has a minimal drop in accuracy.
The drop in recall is more prominent because shallow \eps are unable to
recognize hard-to-detect events, which leads to more false negatives.
In contrast, the drop in precision is minimal.
On \qone, \sys improves both precision and recall.
This is because the \plan augmented with the \me technique overestimates the
\eps for this query, leading to an improvement in accuracy.

\begin{figure}[t]
  \centering
  \includegraphics[width=\linewidth]{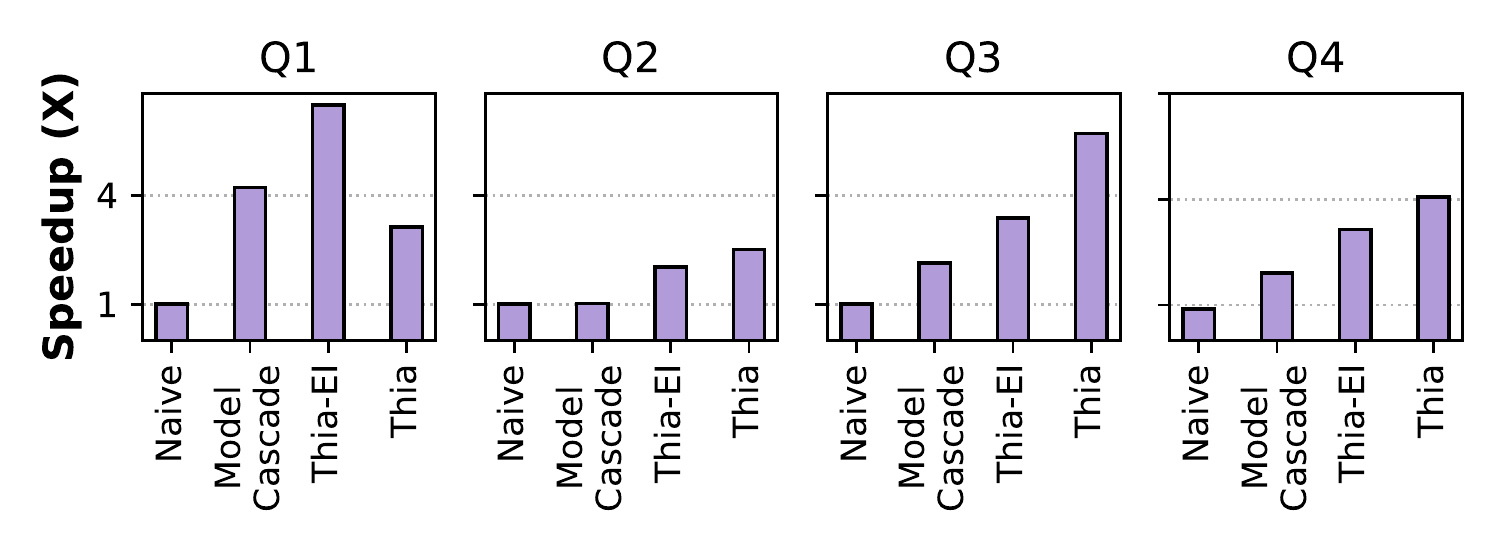}
  \caption{\textbf{Model Cascade \vs \ei} --
  Comparison of the query processing time taken by \sysnaive, \textsc{Model
  Cascade}, \sysei, and \sys.}
  \label{fig:eval:mc_vs_ei}
\end{figure}

\subsection{Model Cascade \vs \ei}~\label{sec:eval:mcopt}
As we mentioned in~\autoref{sec:back:sys}, \sysmulti uses a model
cascade~\cite{tahoma}. 
It differs from the naive model cascade technique in that it uses \ds to
select \eps.
In contrast, a naive approach determines whether to stop at an \ep based on the
a confidence score of the prediction from the previous \ep.
\sysmulti outperforms the naive approach since shallower \eps in the model
cascade are always executed with the latter technique.
As a result, our \ei based systems (\ie \sysei and \sys) outperform the model
cascade approach.
Since it is challenging to construct a confidence-score based system in the
case of object detection, we show the projected performance of the model cascade
approach compared to \sysei and \sys in~\cref{fig:eval:mc_vs_ei}.
The \ei based system delivers 2\X speedup compared to \textsc{model cascade}. 

\begin{figure}
  \includegraphics[width=0.5\columnwidth]{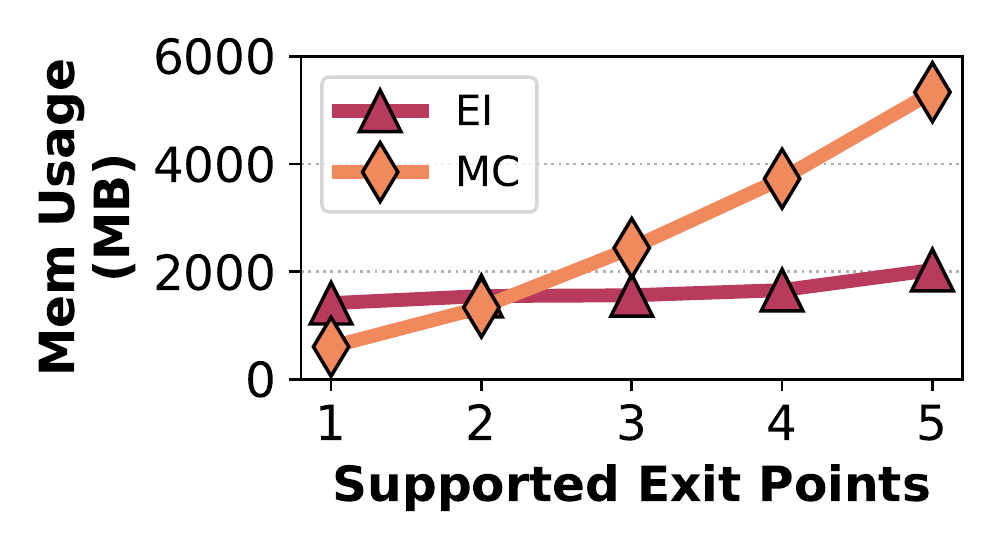}
  \caption{\textbf{Memory Footprint} -- Comparison of memory footprint of
  \sysei and \sysmulti (\ie a model cascade).}
  \label{fig:eval:mem}
\end{figure}

Using multiple models to construct model cascade also increases the memory
footprint of the system.
As shown in~\cref{fig:eval:mem}, the real-time memory usage of a model cascade 
increases when we increase the number of \eps.
It has a 5~GB memory footprint with 5 \eps.
In contrast, since the \ei model shares parameters and inference features, 
it only incurs a 2~GB memory footprint for the same number of \eps.

\begin{figure}[t]
  \centering
  \includegraphics[width=\linewidth]{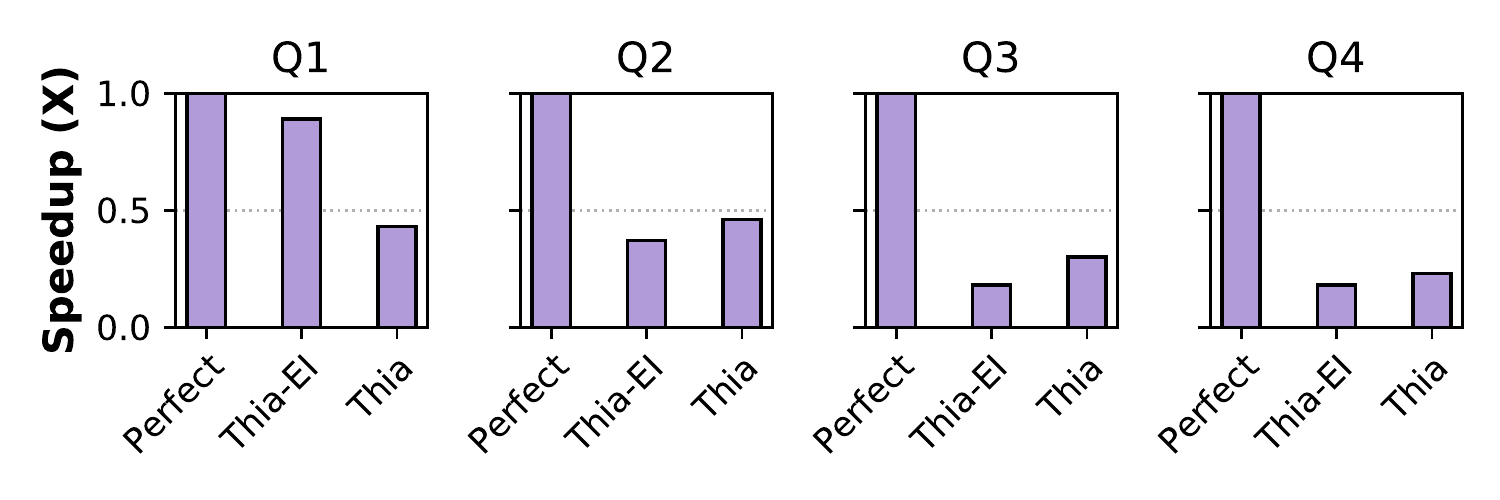}
  \caption{\textbf{Optimality of \ds} -- Comparison of execution time of \sysei
  and \sys against that with the optimal plan.}
  \label{fig:eval:perfect_speedup}
\end{figure}

\subsection{Optimality of \ds}\label{sec:eval:perfect}
We now examine the quality of the query plans relative to the optimal
plan by comparing the execution speedup.
The optimal plan is constructed using a brute-force \ep selection on every
frame (\ie chunk size = 1).
The optimal plan is un-achievable in reality because 
the brute-force selection on every frame significantly increases \sampleover.
The results are shown in~\cref{fig:eval:perfect_speedup}.
The plans constructed by the \plan are 0.3\X slower than the optimal plan.
So, there is still potential for improving the quality of the query plans.
We plan to explore techniques for doing so with a tolerable impact on
\sampleover in the future.

\section{Limitations}
\PP{Accuracy of \me model}.
The \me technique uses a simple neural network to model the optimal \eps
distribution.
However, the inaccuracy of this model leads to a minimal drop in overall query
accuracy.
We plan to study other techniques to reduce the \sampleover in the future.
For example, instead of using a deep learning model, a lightweight statistical
estimator may be sufficient.
A challenge with this approach is that this estimator must accurately map all of
the parameters returned by the object detection model (\eg a set of bounding
boxes and confidence scores) to the appropriate \ep.

\PP{Query Support}.
Currently, \sys supports a limited set of queries. 
To support general-purpose video analytics, we will need to add support for
additional types of queries (\eg aggregate queries).
We plan to integrate the \ei and \ds techniques into the query execution engine
and the query optimizer of a full-featured video analytics system in the
future.
\section{Related Work}
\PP{Model Cascade}. 
Researchers in the area of face detection have proposed models that support a
set of \ep{s} that are geared for different accuracy and speed
trade-offs~\cite{face-cascade-1,face-cascade-2}.
These models return a binary decision and a confidence score (\ie whether a face
exists).
Based on the confidence score, the model chooses the appropriate \ep.
In contrast, \sys uses the estimator to directly pick the \ep.
% \AJ{\ldots}

% %
% Panorama~\cite{panorama} uses a model cascade to solve the unbounded vocabulary
% problem.
% %
% The model contains multiple \eps on the backbone network and uses customized
% output layers tailored for the vocabulary problem.
% %
% Unlike \sys, their model clusters whether two inputs belong to the same
% category and selects the \ep based on the distance between features of two
% inputs.
% %

% Tahoma~\cite{tahoma} uses a cascade of differently-sized models.
% %
% It decides the model to use based on the confidence score of the classification
% decision returned by each model.
% %
% \syspp, NoScope and \sysblaze~\cite{pp, noscope, blazeit} also use a collection
% of specialized models to realize the model cascade idea.
% %
% Unlike \sys, these systems do not dynamically change the model during query
% execution for each chunk.
% %
% They profile all the models on a sample of the dataset, and select the best
% model.

\PP{Query planning}.
The authors of Chameleon~\cite{chameleon} observe that an appropriate query 
plan is critical to gain high performance and accuracy.
Similar to the \ds technique, it adjusts the execution plan at runtime.
To reduce the cost of picking the correct plan, it exploits temporal locality of
nearby frames in the video, thereby reducing the profiling cost.
To further reduce this cost, it uses a clustering algorithm to explore
correlation across videos.
\sys instead uses a shallow neural network to directly estimate the optimal \ep
to use for a chunk.

% \PP{Visual Data System}.
%
% VDMS~\cite{video-db2} differs from \sys in that it does not natively
% support deep learning models.
%
% It instead focuses on the preprocessing phase (\eg image resizing and cropping).
%
%RasDaMan~\cite{video-db3} manages visual data using an intelligent optimizer
% and flexible array-oriented storage engine.
%
%In contrast, \sys stores visual data on disk until needed for query processing.
%

\section{Conclusion}
We presented, \sys, a video analytics system for efficiently processing visual
data at scale.
\sys leverages the early inference technique to support a range of
throughput-accuracy tradeoffs.
It then adopts a fine-grained approach to query planning and processes different
chunks of the video with different exit points to meet the user's requirements.
Lastly, \sys uses a lightweight technique for directly estimating the exit point
using a shallow deep learning model to lower the optimization time.
We empirically show that these techniques enable \sys to outperform two
state-of-the-art video analytics systems by up to 6.5\X, 
while providing accurate results even on queries focusing on hard-to-detect
events.
%
%By automating the detection of infrequently occuring, hard-to-detect objects in
%videos, \sys reduces the labor cost of analysing visual data.

% ref
\clearpage
\bibliographystyle{ACM-Reference-Format}
\bibliography{ref}

\end{document}